%% file: polarized_beams_rep_prog_phys.tex
\documentclass[12pt]{iopart}

\usepackage{iopams}  
\usepackage{graphicx,hyperref}
\usepackage{xcolor}

\usepackage{color}
\usepackage{mathrsfs}
\usepackage{graphicx}
\usepackage{dcolumn}
\usepackage{bm}
\usepackage{hyperref}

\usepackage{epstopdf}
\usepackage{xcolor}
\usepackage{ulem}

\newcommand{\Omegab}{\bm{\Omega}}
\newcommand{\Eb}{\textbf{E}}
\newcommand{\Bb}{\textbf{B}}
\newcommand{\vb}{\textbf{v}}
\newcommand{\ssb}{\textbf{s}}
\newcommand{\Fb}{\textbf{F}}

\newcommand{\D}{\mathrm{d}}

\usepackage{graphicx}
\begin{document}

\title[]{Plasma acceleration of polarized particle beams}

\author{L Reichwein$^{1,2}$, Z Gong$^3$, C Zheng$^{4,5}$, LL Ji$^6$,  A Pukhov$^2$, and M Büscher$^{1,4}$}

\address{$^1$Peter Grünberg Institut (PGI-6), Forschungszentrum Jülich, 52425 Jülich, Germany}
\address{$^2$Institut für Theoretische Physik I, Heinrich-Heine-Universität Düsseldorf, 40225 Düsseldorf, Germany}
\address{$^3$CAS Key Laboratory of Theoretical Physics, Institute of Theoretical Physics, Chinese Academy of Sciences, Beijing 100190, China}
\address{$^4$Institut für Laser- und Plasmaphysik, Heinrich-Heine-Universität Düsseldorf, 40225 Düsseldorf, Germany}
\address{$^5$ExtreMe Matter Institute EMMI, GSI Helmholtzzentrum für
Schwerionenforschung, 64291 Darmstadt, Germany}
\address{$^6$State Key Laboratory of Ultra-intense Laser Science and Technology, Shanghai Institute of Optics and Fine Mechanics, Chinese Academy of Sciences, Shanghai 201800, China}
\ead{l.reichwein@fz-juelich.de; zgong92@itp.ac.cn; chuan.zheng@hhu.de; jill@siom.ac.cn; pukhov@tp1.hhu.de; m.buescher@fz-juelich.de}
\vspace{10pt}
\begin{indented}
\item[]\today
\end{indented}

\begin{abstract}
Spin-polarized particle beams are of interest for applications like deep-inelastic scattering, e.g. to gain further understanding of the proton's nuclear structure. With the advent of high-intensity laser facilities, laser-plasma-based accelerators offer a promising alternative to standard radiofrequency-based accelerators, as they can shorten the required acceleration length significantly. However, in the scope of spin-polarized particles, they bring unique challenges.

This paper reviews the developments in the field of spin-polarized particles, focusing on the interaction of laser pulses and high-energy particle beams with plasma. The relevant scaling laws for spin-dependent effects in laser-plasma interaction, as well as acceleration schemes for polarized leptons, ions, and gamma quanta, are discussed.

\end{abstract}

%\newpage 

\vspace{2pc}
%\noindent{\it Keywords}: spin polarization, plasma-based accelerators, laser-plasma interaction
\submitto{\RPP}
\maketitle

\tableofcontents
\newpage
\section{Introduction}

The study of spin dynamics in plasma has gained significant interest in recent years, specifically in the scope of generating relativistic, spin-polarized particle beams.
Earlier research with polarized particle beams has, e.g., utilized the polarization build-up over time due to the Sokolov-Ternov effect in storage rings or used pre-polarized electron or hadron sources in the case of linear accelerators. A review of spin dynamics in ``conventional'' accelerators was given by Mane in \cite{Mane2005_review_spin}.

However, the developments in high-intensity lasers and the progress of plasma-based accelerators have brought the option of obtaining and accelerating polarized particle beams via laser-plasma interaction within reach. Extensive reviews on the progress in the laser-driven wakefield acceleration of electrons and the different acceleration schemes for ions have been given by Esarey and Macchi \textit{et al.}, respectively \cite{esarey_LWFA_RMP, macchi_2013_RMP}. Reviews geared towards beam-driven wakefield accelerators were given by Hogan \cite{Hogan2016_pwfa_review}, and Adli and Muggli \cite{Adli2016_pwfa_review} in the specific case of a proton driver.

For applications in areas such as high-energy physics and nuclear fusion, where polarization persistence is a requirement, the dynamics of particle spin have to be
thoroughly understood.

In the field of high-energy physics, polarized particle beams can be utilized for deep-inelastic scattering (DIS) to improve the understanding of the proton nuclear structure \cite{Glashausser1979_nuclear} and are of general relevance to quantum chromodynamics \cite{Gross2023_qcd}. 
Unpolarized DIS gives insights into the number density of partons with a momentum fraction $x$ of the hadron. By contrast, polarized DIS, i.e. the collision of a longitudinally polarized lepton beam with a longitudinally or transversely polarized target, yields information on the number density of partons for specific $x$ and spin polarization \cite{Anselmino1995_polarized_DIS}. The spin-dependent scaling functions $g_{1,2}(x)$ are found to have more complicated interpretations in a naive parton model of the nucleons than the unpolarized equivalents $F_{1,2}(x)$. Thus, the polarized process has been used to determine to what percentage the proton spin is carried by the quarks \cite{deFlorian2014_gluon_polarization}. Similarly, polarized Drell-Yan processes can also be used to gain insight into parton distribution functions \cite{Jaffe1991_drell_yan}.

For nuclear fusion, Goldhaber \textit{et al.} proposed in \cite{Goldhaber1934_fusion} that the relevant cross section depends on the relative spin orientation of the reactants. For the reaction $d + t \to \alpha + n$, the enhancement factor is approximately 50\% \cite{Baylor2023_fusion}.
The intermediate step in the reaction, the nucleus ${}^5\mathrm{He}$, has an excited state with angular momentum $J = 3/2$. The spins of the deuteron and triton are $1\hbar$ and $\frac{1}{2}\hbar$, respectively, thus leading to a total angular momentum of $J\hbar$, with $J\in \lbrace \frac{1}{2}, \frac{3}{2}\rbrace$. Since the statistical weight for the $3/2$-state is four, and it is two for the $1/2$-state, random polarization of both reactants yields a probability of $2/3$ that the colliding system ends up in the state of $J = 3/2$ \cite{Kulsrud1986_fusion_calculation}. Thus, if the spins are aligned with respect to the magnetic field, the cross section is increased, yielding a high probability that the excited ${}^5\mathrm{He}$ nucleus is formed, with subsequent decay into $\alpha + n$.
For other reactions, like proton-boron fusion, similar factors have been calculated \cite{Ahmed2013_pB_fusion}.

Low-energy, polarized electrons in the 100 eV range are also of interest for applications in surface physics, including research on spin-orbit coupling and topological materials \cite{Tusche2024_surface_physics}. This has brought forward techniques like spin-polarized low-energy electron diffraction (SPLEED) to study scattering from single-crystal solid surfaces \cite{Kirschner1979_spleed}.

Moreover, the understanding of spin effects in plasma has led to the idea of using polarization as a diagnostic tool for measuring the strength of electromagnetic fields in astrophysical plasma \cite{Gong2021_transient_fields, Gong2023_electron_filament} or for examining the structure of plasma wakefields \cite{An2019_mapping_field}.

The prospects of these applications have led to increased research on sources of polarized particle beams and their plasma-based acceleration. Beyond that, polarimetry tailored to these novel sources has been developed in recent years, as the time structure of the particle beams and the strong electromagnetic pulses induced by the accelerating laser are challenging for conventional detectors.

This review provides an overview of the progress in the acceleration of spin-polarized particle beams via plasma-based acceleration schemes.
The remainder of the introduction discusses the main spin-dependent effects that can become relevant during the interaction of high-intensity laser pulses or high-energy particle beams with plasma.
Section \ref{sec:techniques} describes some of the experimental techniques involved with spin, ranging from the production of polarized particle beams to their detection.
The results on polarized electrons and positrons are presented in section \ref{sec:leptons}. Similarly, theoretical and experimental results on polarized ions are presented in \ref{sec:ions}. The production of polarized gamma quanta is described in \ref{sec:gamma}. Lastly, a summary and a discussion of future prospects are found in section \ref{sec:summary}.

\begin{figure}
    \centering
    \includegraphics[width=0.5\textwidth]{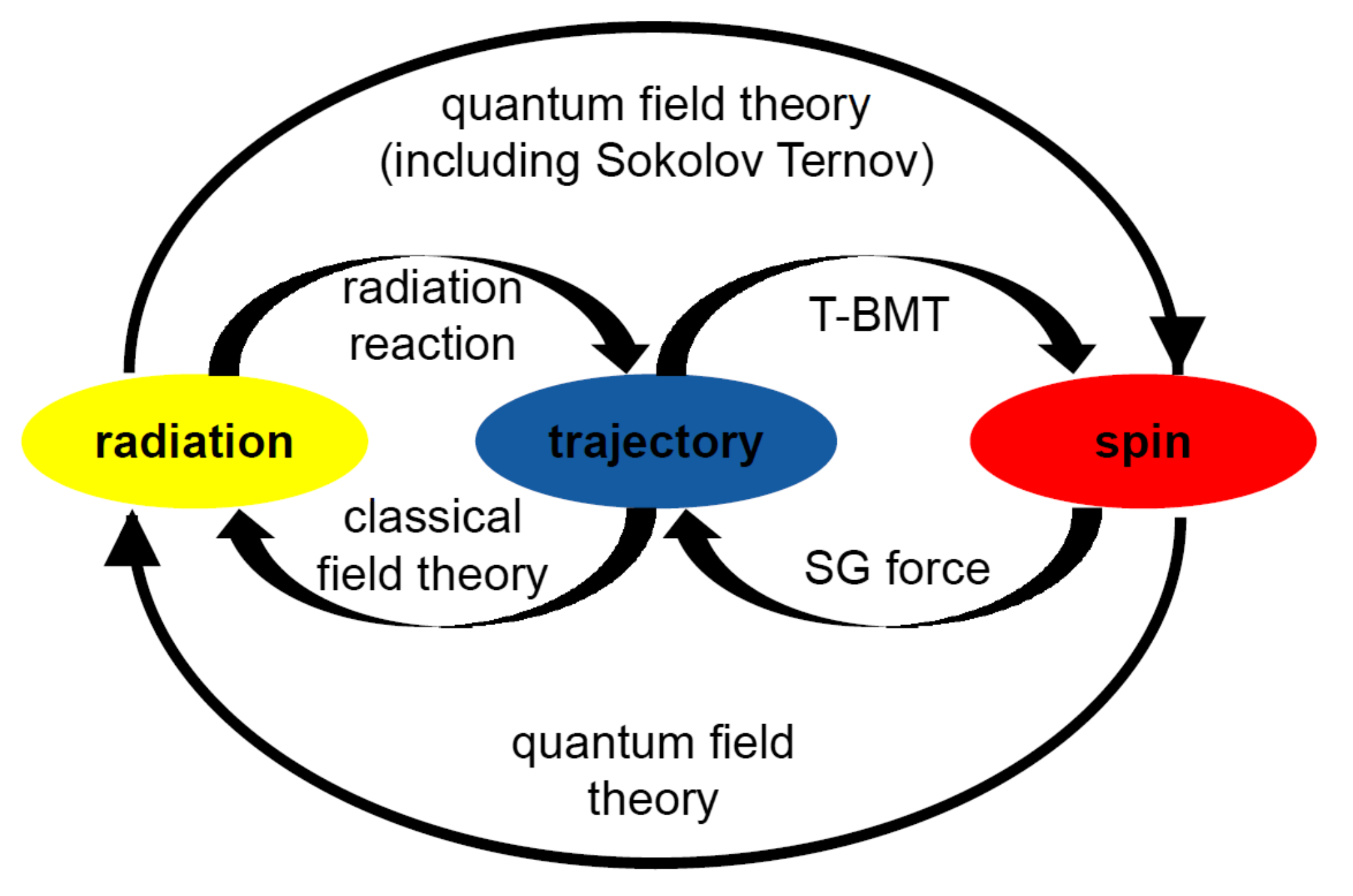}
    \caption{\label{fig:thomas2020_effects} Schematic overview of the relevant effects affecting particle spin in laser-plasma interaction. Reproduced under the terms of the CC-BY license from \cite{Thomas2020_scaling}. Copyright 2020, The Authors, published by American Physical Society.}
\end{figure}

An extensive discussion of the effects relevant to laser-plasma based sources of polarized beams was given by Thomas \textit{et al.} in \cite{Thomas2020_scaling}, in which the main effects discussed are spin precession, spin-dependent trajectories, and radiative polarization. A schematic overview of the connection of the various effects is presented in Figure \ref{fig:thomas2020_effects}.  In the following, we will detail the most relevant aspects of this discussion.

\begin{enumerate}
    \item \textit{Spin precession:} The Thomas-Bargmann-Michel-Telegdi (T-BMT) equation \cite{Bargmann1959_tbmt, Thomas1926_tbmt} describes spin precession according to the equation
\begin{equation}
    \frac{\D \ssb}{\D t} = - \Omegab \times \ssb \; ,
\end{equation}
where

\begin{equation}
    \Omegab = \frac{qe}{mc} \left[ \Omega_B \Bb - \Omega_v \left(\frac{\vb}{c} \cdot \Bb \right) \cdot \frac{\vb}{c} - \Omega_E \frac{\vb}{c} \times \Eb \right]
\end{equation}
is the precession frequency. It depends on the prevalent electromagnetic fields $\Eb, \Bb$ and the particle's velocity $\vb$. Further, $qe$ denotes the charge of the particle, $m$ its mass and $c$ is the vacuum speed of light.
The prefactors
\begin{equation}
    \Omega_B = a + \frac{1}{\gamma} \; , \qquad \Omega_v = \frac{a \gamma}{\gamma + 1} \; ,  \qquad \Omega_E = a + \frac{1}{\gamma + 1} \; , 
\end{equation}
depend on the anomalous magnetic moment $a$ and the Lorentz factor $\gamma$ of the corresponding particle.
The anomalous magnetic moment is defined as $a = (g-2)/2$ via the Land\'{e} $g$-factor. It can be calculated as higher-order loop corrections to the magnetic moment. For the electron it is $a_e \approx \alpha / (2\pi) \approx 10^{-3}$, for protons it is roughly $m_p / m_e$ times larger.
An extensive list of different magnetic dipole moments is given, e.g., by Stone \cite{Stone2005_magnetic_moments}.

Thomas \textit{et al.} derived scaling laws for the precession frequency of electrons and protons in dependence of the dominant field strength $F = \mathrm{max}(|\Eb|, |\Bb|)$. In particular, they found the minimum depolarization time to be
\begin{equation}
    T_{D,p} = \frac{\pi}{6.6 a_e F} \; .
\end{equation}

Within the scope of their paper, the minimum depolarization time is defined as the characteristic time at which the maximum angle between the original direction of the particle spin and the final state is on the order of $\pi / 2$. As discussed later in more detail, particle beams preserve their polarization degree, as long as all particles are subject to the same homogeneous field. If locally varying fields like oscillating laser fields are applied, the polarization is quickly lost due to the differences in precession frequency.

\item \textit{Spin-dependent trajectories:} The most well-known example of spin-dependent trajectories is the Stern-Gerlach effect \cite{Gerlach1922_sg}, in which particles carrying a magnetic moment $\bm{\mu}$ are subject to a gradient magnetic field. The exerted force in that case is
\begin{equation}
    \Fb = \nabla (\bm{\mu} \cdot \Bb) \; .
\end{equation}
This description can be generalized to
\begin{equation}
    \Fb_\mathrm{SG} = \left( \nabla - \frac{\D}{\D t} \nabla_\vb \right) (\Omegab \cdot \ssb) \; .
\end{equation}

For particles in the GeV-range, Thomas \textit{et al.} find the following scaling laws for the maximum particle separation of electrons and protons in terms of the laser wavelength $\lambda_L$ and the relativistic factor $\gamma$:
\begin{eqnarray}
    \Delta_e (\partial F = 0) &\propto 0.3 (2 + 3 a\gamma) \lambda_L [\mu\mathrm{m}]^{-1} \gamma^{-1} \; , \\
    \Delta_p (\partial F = 0) &\propto 0.3 \left( \frac{m_e}{m}\right)^2 \lambda_L [\mu\mathrm{m}]^{-1} \gamma^{-1} \; .
\end{eqnarray}

These scaling laws assume that the prevalent fields are somewhat homogeneous ($\partial F = 0$) like it would be the case in plasma channels.

\item \textit{Radiative polarization:} The Sokolov-Ternov effect describes the build-up of polarization due to spin flips during the emission of radiation \cite{Ternov1995_st}. As there is a slight difference in the probabilities for a flip from spin-up to spin-down compared to the opposite case, a particle beam can gain net polarization over time. This effect is commonly utilized in storage rings.
The polarization of the beam builds up according to
\begin{equation}
    P(t) = P_\mathrm{eq} [1 - \exp(-t/\tau_\mathrm{pol})] \; , \quad  P_\mathrm{eq} = \frac{P_\uparrow - P_\downarrow}{P_\uparrow + P_\downarrow} \; , 
\end{equation}
where $P_\uparrow, P_\downarrow$ denote the different spin-flip probabilities and $\tau_\mathrm{pol}$ is the polarization time.

The scaling laws for the characteristic polarization time in terms of beam energy and field strength found by Thomas \textit{et al.} are
\begin{eqnarray}
    T_\mathrm{pol, electron} &= \frac{10^{-7} \mathrm{s}}{T_e[\mathrm{GeV}]^2 F[\mathrm{TV/m}]^3} \; , \\
    T_\mathrm{pol, proton} &= \frac{10^{14} \mathrm{s}}{T_p[\mathrm{GeV}]^2 F[\mathrm{TV/m}]^3} \; .
\end{eqnarray}

\end{enumerate}

In their paper, Thomas \textit{et al.} conclude that for most laser-plasma based accelerators (not including future high-intensity laser facilities), spin precession according to the T-BMT equation is the only effect of concern for polarization preservation, while the Stern-Gerlach force and the Sokolov-Ternov effect can mostly be neglected.

More extensive discussions on radiative spin dynamics have been given by several authors in the framework of quantum electrodynamics (QED), like studies on spin-dependent radiation reaction by Seipt \textit{et al.} \cite{Seipt_2019}. Geng \textit{et al.} identified a spin-dependent deflection mechanism based on radiation reaction that occurs during the laser-electron interaction \cite{Geng2020_radiative_effect}. 
The effect is shown to dominate the Stern-Gerlach force for electrons with 100s of MeV and 10 PW-class laser pulses.
Over the years, various models describing radiative polarization have been proposed \cite{Seipt2018_radiative, Guo2020_radiative, Song2021_radiative, Tang2021_radiative_dynamics, Artemenko2023_radiative, qian2023parametric}, which are discussed in more detail in the sections concerning polarized positrons and gamma quanta.
A kinetic description of relativistic spin-polarized plasma was derived by Seipt and Thomas in \cite{Seipt2023_kinetic_theory}. Therein, QED effects are taken into account via Boltzmann-type collision operators and the approximation of locally constant fields.

The various effects discussed here, in particular the precession according to T-BMT and radiative polarization in the context of QED, have been incorporated into various particle-in-cell (PIC) codes \cite{Brodin2013_spin_pic, Crouseilles2021_spin_pic}. Moreover, nine-dimensional particle pushers for an accurate description for relativistic particles in strong fields have been proposed, e.g. by Li \textit{et al.} \cite{Li2021_9d}.

\section{Experimental techniques} \label{sec:techniques}

In this section, we will cover aspects of targetry and polarimetry, in particular, how polarized targets for laser-plasma interaction may be produced and how the degree of polarization can be measured.

\subsection{Targets}
For the experimental realization of polarized beam generation
from laser-induced plasmas, the choice of the target
is the most crucial point. Several concepts can be found in literature, split into (i) pre-polarized targets, where the spins of the particles to be accelerated are already aligned before the acceleration process sets in, and (ii) unpolarized targets, which require a spin-selective ionization or injection during the acceleration process.

\subsubsection{Pre-polarized targets}

\begin{figure}
        \centering
        \includegraphics[width=0.75\textwidth]{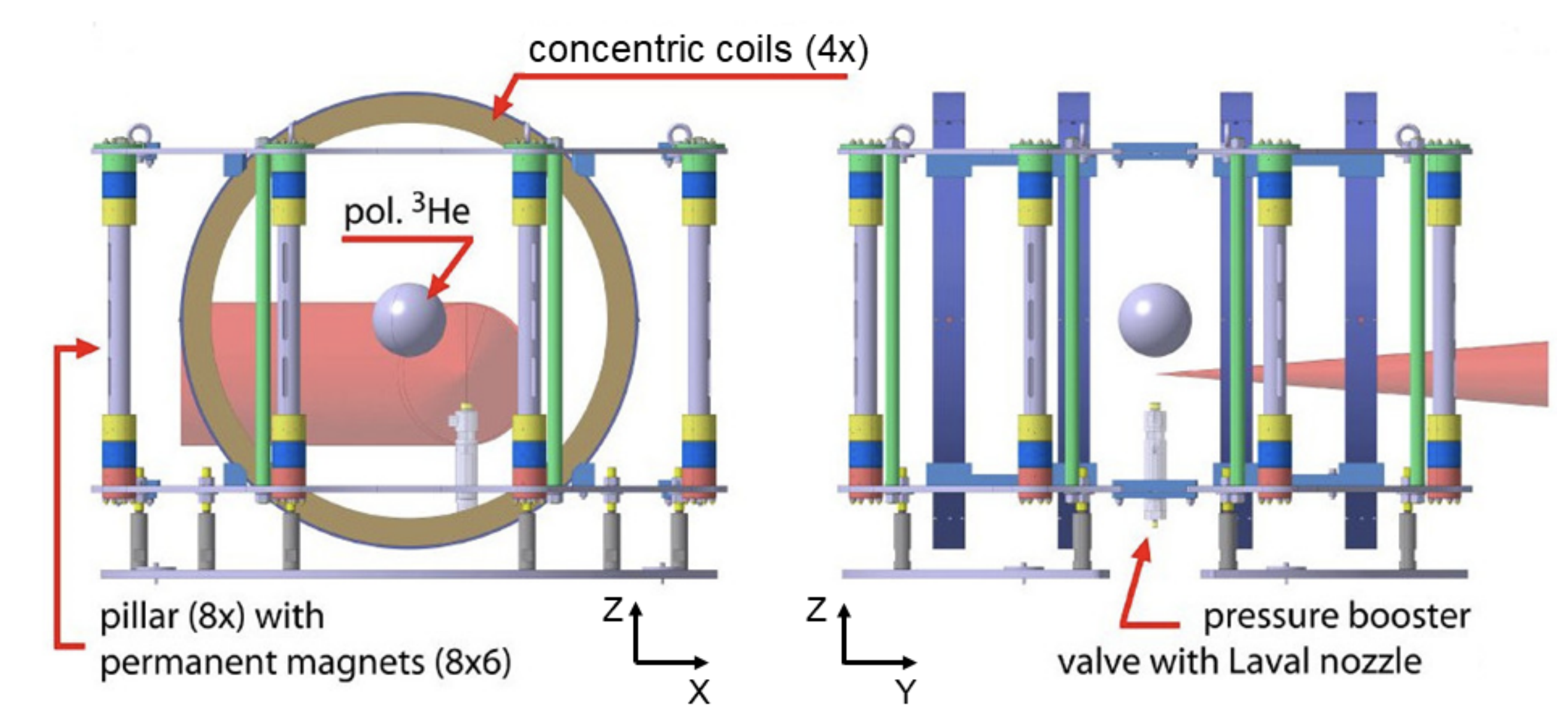}
        \caption{\label{fig:fedorets_he3}Magnetic holding setup for the acceleration of polarized Helium-3. Reproduced under the terms of the CC-BY license from \cite{Fedorets2022_he3}. Copyright 2022, The Authors, published by MDPI AG.}
    \end{figure}

The only polarized target that has been used so far for laser-plasma acceleration experimentally employs a jet containing nuclear polarized $^3$He gas \cite{Fedorets2022_he3}. This rare helium isotope has the advantage that it can be rather easily polarized at room temperature and be stored for many hours in a magnetic holding field (cf. Fig. \ref{fig:fedorets_he3}). High degrees of nuclear polarization ($P\sim 40$\%) can be obtained in a gas jet at densities of a few times $10^{19}$\,cm$^{-3}$. The polarization for this and similar targets can be achieved via methods like metastability-exchange optical pumping \cite{Mrozik2011_meop} or spin-exchange optical pumping \cite{Anderson2020_seop}.
Data of an experiment at the PHELIX laser suggest persistence of this polarization after acceleration of the $^3$He ions to MeV energies \cite{Zheng2023_evidence_he3}. A disadvantage of pre-polarized targets, such as $^3$He gas, is that a permanent magnetic holding field must be provided around the interaction zone in order to maintain the target polarization. In the case of $^3$He gas, the field strength requirement is rather low, on the order of a few mT, but great care has to be taken with the homogeneity of the holding field \cite{Fedorets2022_he3}. Such fields can either be provided by permanent-magnet assemblies or by Helmholtz-type coils. The latter have the disadvantage of requiring cooling (if mounted inside the laser-interaction chamber), but allow the initial $^3\mkern-2mu$He polarization to be rotated relative to the laser propagation axis, thus providing maximum flexibility in adjusting the beam polarization. 

Gas targets containing high-density spin-polarized hydrogen (SPH) have been proposed for the acceleration of electrons (see Ref. \cite{Popp2021,sofikitis2024highenergypolarizedelectronbeams} and references therein) as well as proton or deuteron beams (see for example Ref. \cite{Jin2020_mva}). The SPH is produced through the photo-dissociation of unpolarized hydrogen halides and contains --- dependent on the application --- either electrons or protons/deuterons with high degrees of polarization. The principle behind these targets is to first align the molecular bond of diatomic gas molecules in the electric field of a linearly polarized laser pulse. This is followed by the photo-dissociation into molecular states with polarized valence electrons using a second, circularly polarized UV laser pulse. The hyperfine structure of hydrogen atoms then causes the polarization to oscillate from the electron to the proton (deuteron) and backwards, with a period of 0.7\,ns. Finally, a third laser pulse drives the acceleration process. The advantage of these targets is that the polarized target material can be resupplied to the interaction zone at high rates, no magnetic holding fields are required and densities close to critical density can be achieved.

\begin{figure}
        \centering
        \includegraphics[width=0.75\textwidth]{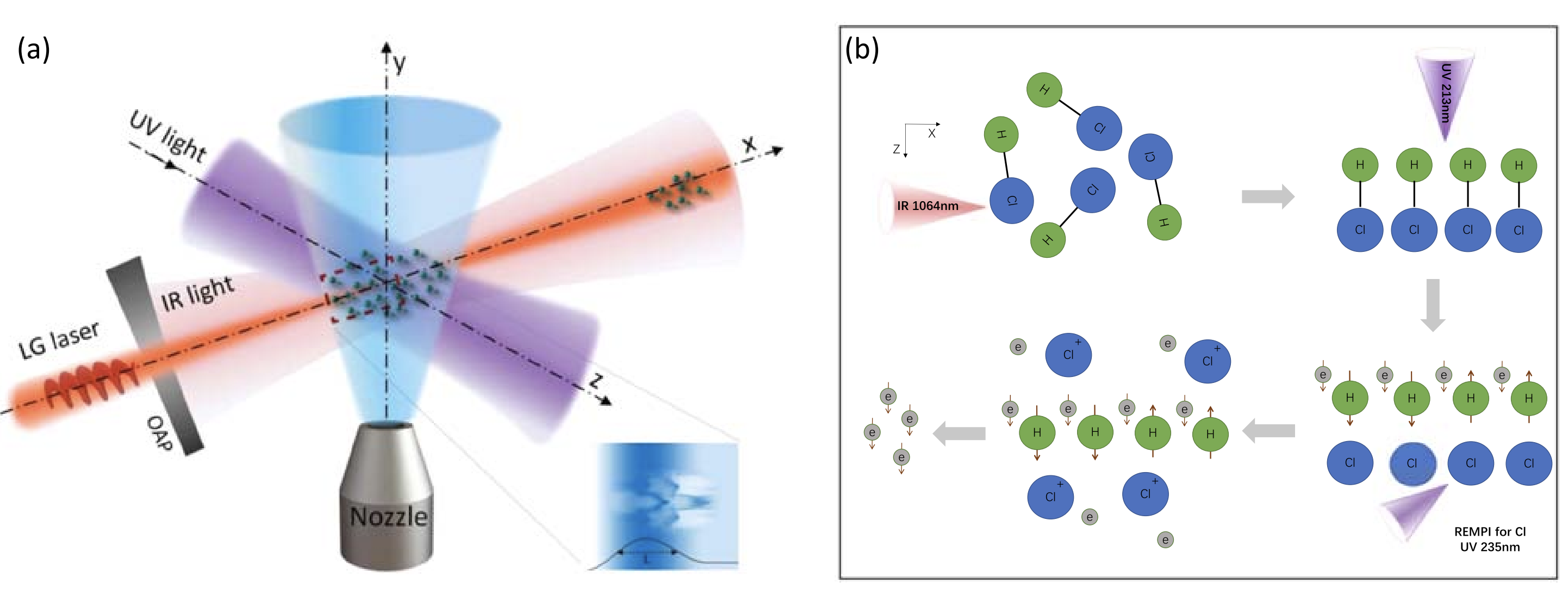}
        \caption{\label{fig:wu2019_hcl}(a) Potential experimental configuration for laser-driven wakefield acceleration of spin-polarized electrons. (b) Schematic of polarized electron production from an HCl target using multiple lasers for alignment, photo dissociation and ionization. Reproduced under the terms of the CC-BY license from \cite{Wu2019_lwfa}. Copyright 2019, The Authors, published by IOP Publishing Ltd.}
    \end{figure}

While the  use of SPH targets seems rather straightforward for proton acceleration \cite{Jin2020_mva,Huetzen2019_protons}, the acceleration of polarized electrons would first require the removal of unpolarized electrons from the accelerating plasma which are unavoidably present in the (otherwise unused) halide atoms. Two solutions have been put forward here, either removal via ionization \cite{Wu2019_lwfa} or spatial separation of the SPH from the unwanted photoproducts leaving behind only hydrogen atoms of high polarization ($P = 90$\%) \cite{sofikitis2024highenergypolarizedelectronbeams}. It remains to be seen which of these approaches will ultimately prove to be more practicable. A schematic detailing the production of polarized electrons from an HCl target is presented in Fig. \ref{fig:wu2019_hcl}. 

Kannis and Rakitzis have proposed an alternative using formaldehyde or formic acid \cite{Kannis2021_molecules}. These molecules can be excited using infrared laser light. The gained rotational polarization can be subsequently transferred to nuclear polarization via hyperfine polarization beating. Afterwards, the molecules are photodissociated to the target molecules $\mathrm{H}_2$ or $\mathrm{H}_2\mathrm{O}$. The production rates of this mechanism can approach $10^{21}\; \mathrm{s}^{-1}$, in contrast to molecular-beam separation techniques which deliver rates of only up to $10^{12}\; \mathrm{s}^{-1}$. This is of importance of potential applications in nuclear magnetic resonance and polarized fusion tests, which require rates of at least $10^{17}\; \mathrm{s}^{-1}$.

According to Ref. \cite{Engels2020molecules}, hyperpolarized cryogenic foils can be produced after recombination of nuclear polarized hydrogen atoms from an atomic beam source to hydrogen-deuterium molecules. These could in principle be used as targets for plasma acceleration of polarized proton or deuterium beams. Due to their technically demanding realization \cite{Engels:2019Gi}, it has yet to be shown that this type of target is suitable for laser-plasma applications.
A theoretical study of the polarization dynamics in the hydrogen isotope molecules, H$_2$, D$_2$, and HD can be found in Ref. \cite{Kannis2018_pol_HD_molecules}. It could be shown that in the presence of a static magnetic field the molecular rotation polarization can be transferred and maintained in the nuclear spin, providing highly nuclear-spin-polarized molecules ($P>90$\%). 

Preliminary data from an experiment with metastable hydrogen atoms indicate that large nuclear polarizations can be produced in a so-called Sona transition unit \cite{Engels2023_hyperpol}. It remains to be seen whether the technique can also be applied to ground-state atoms or molecules and thus be useful to produce polarized targets for laser-plasma acceleration.

\subsubsection{Unpolarized targets}

An unpolarized foil target was used for the first polarization measurement of laser-accelerated particles. The authors of Ref.~\cite{Raab2014_pol_measurement} report on the TNSA acceleration of few-MeV protons from the rear surface of a gold foil. Since no polarization build-up of the protons during the acceleration was observed, it was concluded that pre-polarized targets would be needed to produce polarized hadron beams. 

In Ref. \cite{Nie2021_in_situ}, the use of a gas target containing a mixture of lithium and xenon atoms is proposed. It is predicted that polarized electron beams can be accelerated in a beam-driven plasma wakefield by in-situ injection of polarized electrons from the xenon atoms into the plasma wake generated from the lithium atoms. Similarly, a single-species target consisting of only ytterbium has been proposed \cite{Nie2022_single_species}. This method requires the targeted ionization of specific orbitals with a circularly polarized laser pulse to achieve polarization.

\subsection{Polarimetry}
In order to be experimentally accessible, the information about beam polarization has to be converted into measurable quantities. 
This is generally achieved by sampling the spin projection of a large
number of particles in a secondary scattering process that is sensitive to the spin direction. Both the scattering target and the particle detectors are typically combined in a single device, the so-called polarimeter. Such polarimeters are frequently  used at conventional particle accelerators to measure beam polarizations or polarization-dependent scattering cross sections. However, at laser-plasma accelerators, particular challenges have to be overcome, such as the time structure of the beams and strong electromagnetic pulses induced by the accelerating laser. This often prohibits the use of sophisticated electronic detectors to resolve individual beam particles. Since the scattering processes in polarimeters  
crucially depend on the particle species and energy these must be carefully customized for the respective application.

The first polarimeter for laser-accelerated particles was used to measure the polarization of TNSA-accelerated, few-MeV protons at the 
100 TW Arcturus laser facility, Düsseldorf \cite{Raab2014_pol_measurement}. The polarization measurement is based on the spin dependence of elastic proton scattering off nuclei in a silicon target. This scattering process has a non-vanishing analyzing power $A_y$ which is a measure for the asymmetry of the scattered protons' angular distribution along the azimuthal angle for a given transversal beam polarization. Silicon is well suited as scattering target material at proton energies of a few MeV since high precision data for $A_y$ are available from literature. The scattered protons were recorded behind the scattering target with CR-39 detectors for each individual laser shot. The authors of Ref. \cite{Raab2014_pol_measurement} conclude that  polarizations down to $P\sim20$\% can be detected for particle bunches with a sufficiently high number of produced protons ($\sim 10^8$ with energies above 2.5\,MeV). 

A compact polarimeter for $^3$He ions with special emphasis on the analysis of short-pulsed beams accelerated during laser–plasma interactions is described in Ref. \cite{Zheng2022_polarimetry_he3}. The  polarimeter was tested at a tandem accelerator with unpolarized $^3$He beams and then used for the analysis of polarized laser-accelerated ions \cite{Zheng2023_evidence_he3}. It is based on the detection of $\alpha$ particles in CR-39 stacks, emitted from the $d$-$^3$He fusion reaction with non-vanishing $A_y$ in a deuterated-polyethylene foil. The polarimeter is especially useful for the analysis of $^3$He ion beams in the few-MeV range, but also at lower energies down to 0.6 MeV. The accuracy of a beam polarization measurement, i.e., the minimal detectable $^3$He beam polarization, is estimated to be 21\% for a typical laser–plasma experiment where roughly $10^9$ $^3$He ions enter the polarimeter.

An electron polarimeter for the LEAP project is being prepared at DESY, Hamburg \cite{Popp:2024Cz}. For their expected electron beam energies of tens of MeV, Compton transmission polarimetry is the ideal method to measure the polarization. $\gamma$ photons produced by bremsstrahlung are transmitted through an iron absorber magnetized by a surrounding solenoid, with rate and energy spectra depending on the relative orientation of the photon spin and the magnetization direction. The transmission asymmetry with respect to the magnetization direction  is proportional to the initial electron polarization. Calibration measurements with unpolarized beams are in progress. At higher electron energies (100 MeV and above), the asymmetries for Compton transmission polarimetry decrease and this method becomes impractical. Mott-, M{\o}ller- or laser-Compton-polarimetry must be used instead, however, such polarimeters have not yet been developed for laser-plasma applications.

Laser-Compton polarimetry usually refers to Compton back-scattering  polarimeters~\cite{Beckmann2002}, measuring a longitudinally polarized electron beam at high energies ($\gtrsim1$\,GeV) via the collision with circularly polarized laser light (532\,nm, photon energy 2.33\,eV), which produces backscattered photons ($\gtrsim10$ MeV)~\cite{Gaskell2023_chapter} with a rate depending on the the spin-dependent Compton cross section. The applicability of this method to electron beams from laser-driven wakefield acceleration (LWFA) is very limited, since it would require a precisely calibrated beam-intensity monitor to be operated together with the Compton polarimeter. A non-destructive diagnostic method for LWFA electron bunches based on Transition Radiation (TR) is being  studied \cite{Downer2018}, but the technique is still under development~\cite{Bajlekov2013, Heigoldt2015} with a comparison to the standard destructive method for energy and charge measurement by a magnetic spectrometer.

Compton transmission polarimetry has been performed for a polarized electron beam of 3.5\,MeV at the MAMI injector \cite{Barday2011}, a longitudinally polarized electron beam ($P_{e}\sim 80\%$) at the A4 MAMI experiment at 570\,MeV and 854\,MeV \cite{Weinrich2005}, and also as relative electron polarization monitors at 20 and 200\,MeV at MIT Bates~\cite{Zwart2003}, or for polarized positrons (selected energies from 4.5 to 7.5\,MeV) and electrons (6.7\,MeV) with undulator-based production from E166 at SLAC~\cite{Alexander2009}. A recent application is for DC and SRF electron photo-injectors at JLab and BNL, i.e., spin-polarized electron beam sources at 5 and 7\, MeV \cite{Blume2024}. Compton transmission polarimeters have several advantages for LWFA electron bunches. First, they can be applied for electron (or positron and $\gamma$-photon) bunches with any time structure \cite{Fukuda2003}, which is crucial, since the duration of the bunches is generally less than 10\,fs and the number of particles is about $10^9$ ($\sim$0.1\,nC, and the equivalent current $\sim$10\,kA), ruling out alternative methods. Second, the Compton transmission polarimeters can be used for non-destructive measurements at high energies since the electrons only loose a small fraction of their energy (below a few MeV, also keeping most of their polarization) after a thin radiator and can be separated by a sweeping magnet~\cite{Zwart2003}. This method allows measuring the beam polarization and the beam energy (with an additional electron spectrometer) simultaneously. The beam profile and energy distribution have shot-to-shot fluctuations (laser-plasma instabilities~\cite{esarey_LWFA_RMP, Joshi2020}) in LWFA, so the correction must be included in the polarization measurement. Finally, it is suitable for very high beam intensities because the transmitted $\gamma$-rays are collimated to a small forward cone and reduced by the magnetized absorber and end in a calorimeter with a large energy capacity. 
As discussed, several of the aforementioned polarimetry techniques have been successfully demonstrated. However, adapting them to the developing array of sources for leptons, ions, and gamma quanta described in the subsequent chapters remains an on-going challenge.

\section{Polarized electrons and positrons} \label{sec:leptons}

Within this section, we transition to the theoretical results obtained in studies of the acceleration of polarized lepton beams. Most of the results for electrons are concerned with wakefield acceleration. As mentioned in the previous section, most of the following setups are based on pre-polarized targets, and only a few make use of ``single-step'' schemes to polarize and directly accelerate the particles.

\subsection{Electrons}

  A first investigation of polarized electron beams in wakefield acceleration was conducted by Vieira \textit{et al.} in \cite{Vieira2011_spin_wakefield}. They showed that lower depolarization is observed for the high-energy stages of the accelerator. Analytical expressions for the depolarization of zero-emittance beams were derived. Further, it was shown that externally guided propagation is beneficial to polarization in comparison with self-guiding. The models are in good agreement with PIC simulations. 

  Further studies on spin precession in wakefields were performed by Pugacheva \textit{et al.} in \cite{Pugacheva2016_tbmt}. It was shown that the minimal depolarization occurs for electrons that are injected close to the maximum of the accelerating force with the wake's phase velocity. In a follow-up publication \cite{Pugacheva2018_synchrotron}, the effects of synchrotron radiation on spin dynamics were discussed.

  Wu \textit{et al.} and Wen \textit{et al.} extensively investigated the acceleration of polarized electrons in laser-driven wakefields (LWFA) by 3D-PIC simulations in \cite{Wu2019_lwfa, Wen2019_wakefield} for the first time. The simulations of Wu base their investigation on an HCl target in a density regime of $10^{18}$ cm$^{-3}$. It was shown that for conventional Gaussian laser pulses, the strong azimuthal magnetic field in the induced wake structure is detrimental to the beam polarization. The results could be improved by utilizing Laguerre-Gaussian modes, which exhibit significantly weaker azimuthal fields during injection, leading to approx. 80\% polarization. It was further concluded that the most significant changes in beam polarization occur during injection, while acceleration in regimes where $\gamma \gg 1$ only contributes marginally, since the spin precession according to the T-BMT equation is dependent on the Lorentz factor. In a separate publication, Wu \textit{et al.} have investigated the azimuthal dependence of beam polarization in a wakefield and propose the use of an x-shaped slit as a means of improving the polarization degree \cite{Wu2020_filter}. The simulations indicate that beam polarization may be increased significantly (in their specific case from 35\% to approx. 80\%) with this mechanism.

  The influence of bubble geometry on spin dynamics was investigated by Fan \textit{et al.} \cite{Fan2022_bubble_geometry}. Spherical bubbles are shown to preserve polarization better than aspherical ones, where the net polarization direction can even reverse for oblate bubble shapes.

Yin \textit{et al.} further investigated self-injection for LWFA in \cite{Yin2024_self-injection}. In particular, it is found that for longitudinal self-injection, that the electrons move around the laser axis and the net influence of the laser pulse as well as the contribution of the wakefield can be ignored. Accordingly, ultra-short electron beams with 99\% polarization can be generated using longitudinal injection.

\begin{figure}
    \centering
    \includegraphics[width=0.65\textwidth]{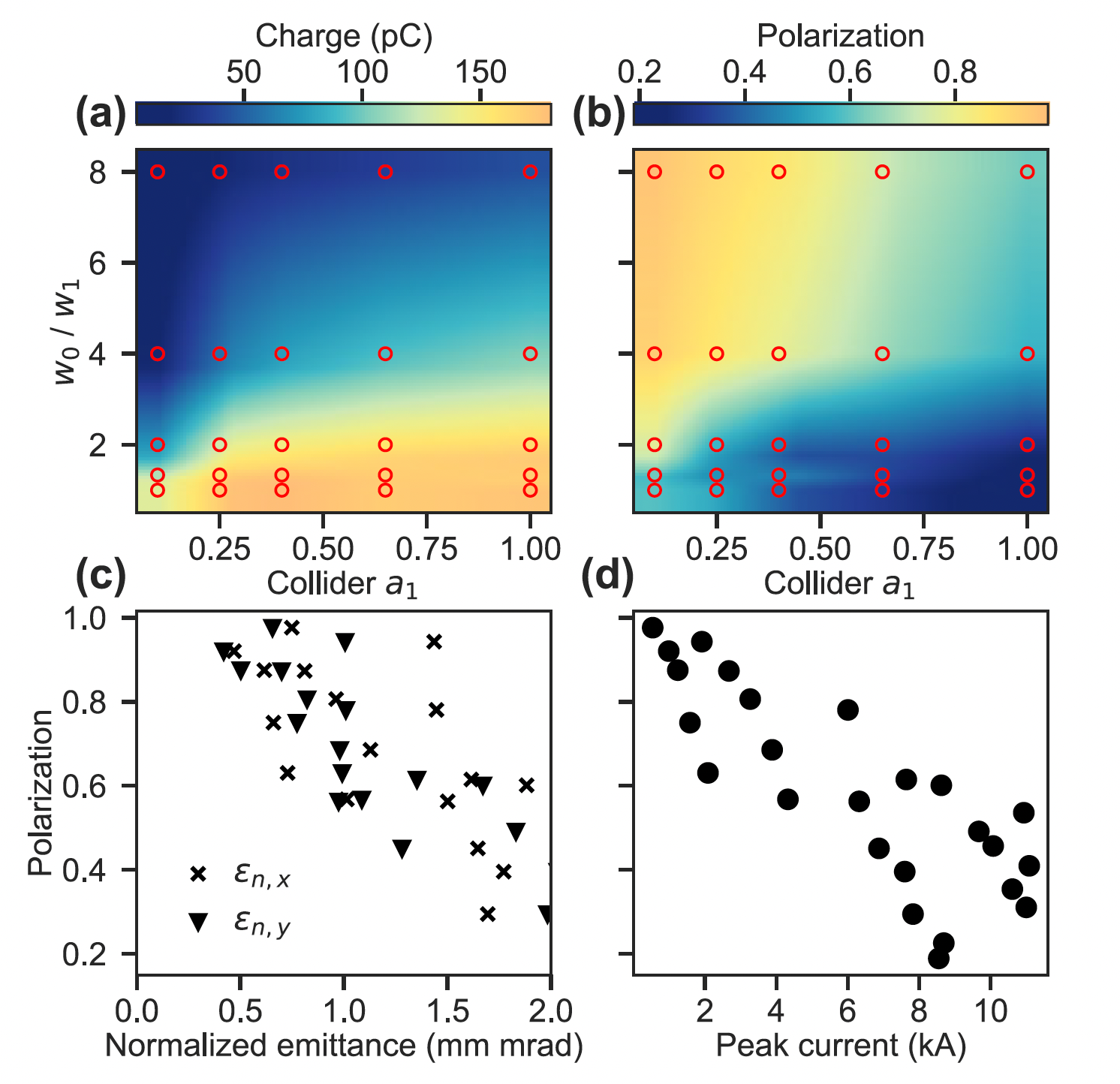}
    \caption{\label{fig:bohlen_colliding} Dependence of beam charge and polarization degree, as well as emittance and peak current, on parameters of the colliding laser pulse. The driving laser pulse is fixed to $a_0 = 2.5$ and $w_0 = 20$ {$\mu$}m. Reproduced under the terms of the CC-BY license from \cite{Bohlen2023_colliding_pulse}. Copyright 2023, The Authors, published by American Physical Society.}
\end{figure}

An alternative means of injection has been discussed in the form of colliding-pulse injection. Gong \textit{et al.} proposed that the driving laser pulse collides with a weaker laser pulse, which leads to a heat-up of electrons in the region of their interaction \cite{Gong2023_colliding_pulse}. Due to this heating, electrons can be injected into the wake. The mechanism is examined for the polarized HCl target at $10^{18}$ cm$^{-3}$. In \cite{Bohlen2023_colliding_pulse}, the driving pulse is fixed at $a_0 = 2.5$ and $w_0 = 20$ {$\mu$}m. The intensity and focal spot size of the laser pulse are varied, examining the influence on beam charge and polarization. It is found that similar spot size between driving and colliding pulse, as well as higher pulse intensity lead to an increase in charge, however at the cost of polarization (cmp. Fig. \ref{fig:bohlen_colliding}). Electron beams with 80\% polarization with tens of pC can be generated using 10 TW-class laser systems; beams in excess of 90\% polarization are achievable for lower charge.

Sun \textit{et al.} proposed a dual-wake injection scheme \cite{Sun2024_dual_wake}, in which a tightly focused, radially polarized laser pulses propagates through a plasma with an abrupt density jump. This mechanism can be used to drive dual wakes in the quasi-blowout regime and produces electron beams as short as 500 attoseconds with a polarization exceeding 80\%. 

The case of beam-driven wakefields was first discussed by Wu \textit{et al.} in \cite{Wu2019_pwfa}. Therein, the influence of driving beam parameters on the witness are studied. Analytical scaling laws for beam polarization in dependence of the peak current are derived, which show that polarization decreases for increases flux. Up to 7.5 kA of electrons with 80\% polarization are obtained from simulations, which is comparable to their LWFA results with respect to the polarization degree.
Beam-driven wakefields can have several advantages over LWFA as effects like laser diffraction and dephasing do not occur. However, beam-driven approaches require a particle source which is not readily available when compared to laser sources, and other detrimental effects like the hosing instability have to be mitigated depending on the specific parameter regime.

For the approaches using pre-polarized targets in connection wakefield acceleration, it should be noted that many publications consider a background density in the range of $10^{18} \;\mathrm{cm}^{-3}$, whereas experimentally feasible densities are currently in the regime of $\leq 10^{17} \;\mathrm{cm}^{-3}$ \cite{sofikitis2024highenergypolarizedelectronbeams}.
Moreover, the volume containing polarized particles that is available for laser-plasma interaction is limited, restricting the injection/acceleration methods for polarized electrons significantly. This is due to method with which the hydrogen halide targets are prepared: the second dissociation laser pulse is responsible for shaping the volume in which only the polarized hydrogen is located. Since collisions which would otherwise decrease the polarization degree have to be avoided, the maximum attainable density is limited. In the case of \cite{sofikitis2024highenergypolarizedelectronbeams}, the HCl target is modeled as a channel of 20 {\textmu}m width and a density of $10^{17} \;\mathrm{cm}^{-3}$. The laser pulse driving the acceleration process has a wavelength of 1.6 {\textmu}m and $a_0 = 6$. An electron beam with 3.9 pC can be produced that way, maintaining 48\% of its initial polarization. As the available degree of pre-polarization depends on channel width and target density (here: 45 \%), this means that effectively only 22\% polarization can be achieved based on these target restrictions.

\begin{figure}
    \centering
    \includegraphics[width=\textwidth]{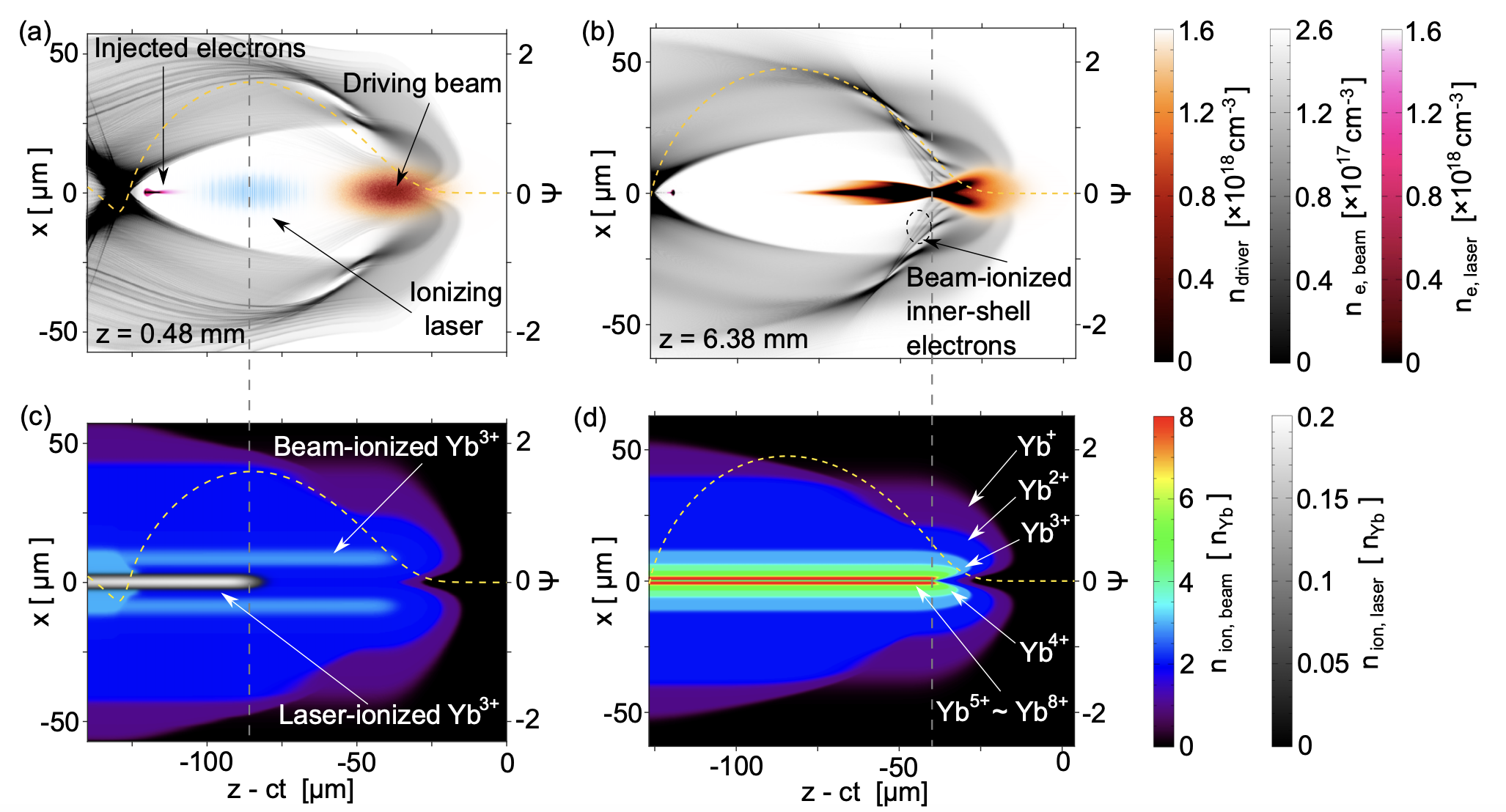}
    \caption{\label{fig:nie_photocathode} The single-species photocathode scheme for polarized electrons. (a), (b) show the plasma and beam density around the laser focus and after driver pinching, respectively. (c), (d) show the Yb ion density at the corresponding time steps. Reproduced under the terms of the CC-BY license from \cite{Nie2022_single_species}. Copyright 2022, The Authors, published by American Physical Society.}
\end{figure}

The aforementioned studies for laser- and beam-driven wakefields generally assume a pre-polarized target, meaning that target preparation and acceleration is a two-step process. Nie \textit{et al.}, however, proposed methods to achieve polarized electron beams \textit{in-situ} \cite{Nie2021_in_situ}. The electrons are injected into the wakefield and simultaneously polarized via the ionization of the outermost $p$-orbital of the utilized noble gas. In the publication, a target consisting of lithium and xenon is used. An electron drive beam ionizes only the lithium and generates the wakefield. A circularly polarized laser pulse is then used to strong-field ionize the $5 p^6$ electron of the xenon atoms. These electrons are spin-polarized and can be accelerated to approx. 2.7 GeV in 11 cm. A net polarization of about 31\% is achieved with this setup.
In a subsequent publication, Nie \textit{et al.} investigate the potential of single-species (ytterbium) plasma photocathodes for polarized beams \cite{Nie2022_single_species}. Here, the electron beam predominantly ionizes the  outer $6s$ electrons and drives the wakefield. A circularly polarized (CP) laser pulse is then used to inject the $4f^{14}$ electrons which is shown to deliver 56\% net polarization with 15 GeV in 41 cm. The setup is shown in Figure \ref{fig:nie_photocathode}.

Beyond the concept of wakefield acceleration from pre- or unpolarized targets in the setting of classical physics, other studies have focused on the radiative polarization dynamics of electrons in intense electromagnetic fields, like Tang \textit{et al.} who propose a self-consistent model that extends the applicability of the polarization vector model to electromagnetic field structures relevant to laser-plasma interaction \cite{Tang2021_radiative_dynamics}. Zhu \textit{et al.} follow an approach utilizing beam-solid interaction \cite{Zhu2024_electrons_target}: here, a relativistic, unpolarized electron beam is incident onto a solid target at a grazing angle. This leads to strong magnetic fields and, in turn, strong beam self-focusing via magnetic pinching. The magnetic fields reach the level of a few Giga-Gauss, triggering synchrotron radiation and accordingly, radiative spin flips. In their simulations, they find that an initial 5 GeV electron beam can generate 50\% polarization for electrons with $< 2$ GeV using their approach. On a similar note, Xue \textit{et al.} utilize a double-layer solid target which is ionized by the incident electron beam \cite{Xue2025_electrons}. The ionization leads to an asymmetric plasma field which can induce spontaneous radiative polarization. This method has the added benefit that the generated fields can further focus the polarized beam. In their 3D-PIC simulations, they are able to produce pC-level electron beams with a polarization of approx. 57\%.
While these approaches provide high polarization and energy of the resulting electron beams, they also require near-future laser facilities for realization.
In contrast, Seipt \textit{et al.} proposed the polarization of high-energy electron beams using pulsed bi-chromatic laser fields, which could be realized with current-day parameters, but this delivers a lower beam polarization of approx. 17\% \cite{Seipt_2019}.

As a potential application, Gong \textit{et al.} showed that initially unpolarized targets of near-critical density can be used as a diagnostic tool for transient magnetic fields \cite{Gong2021_transient_fields}. They find that the angle-resolved electron spin polarization can be used to be predict the leading order of the quasi-static magnetic field induced by the longitudinal current. The method is shown to be robust to moderate imperfections of both laser pulse and plasma density profile.
Similarly, An \textit{et al.} proposed to utilize polarized electron beams to reconstruct the structure of a wakefield, providing advantages compared to conventional charged particle beam probes for more complex structures \cite{An2019_mapping_field}. Further, the radiative polarization of electrons during current filamentation instability has been shown to offer a novel pathway to study astrophysical plasma, allowing to differentiate between different instability regimes like (ab-)normal filamentation and quenching \cite{Gong2023_electron_filament}.
Further references on radiative polarization can be found in the introductory part of this review as well as the section concerning polarized gamma quanta.

\subsection{Positrons}

The work on polarized positron beams is currently divided between the production and the acceleration of these beams. In contrast to electrons, no pre-polarized targets of positrons can be produced, therefore production via processes like Breit-Wheeler pair creation is necessary. This leaves the realization and verification of the proposed mechanisms up to near-future facilities. Moreover, the efficient acceleration of positrons utilizing wakefields (mostly without considering spin) is still subject of current research, as detailed below.

Chen \textit{et al.} proposed the use of two-color laser pulses for the generation of polarized positron beams \cite{Chen2019_two_color}. It is shown that the spin asymmetry in strong external fields together with the asymmetric two-color laser field leads to the creation of beams with up to 60\% polarization. 

An alternative method is described by Li \textit{et al.}: when a high-intensity CP laser pulses interacts with a longitudinally polarized, relativistic electron beam, a highly polarized positron beam can be created via the non-linear Breit-Wheeler process \cite{Li2020_helicity_transfer}. The polarization is transferred from the electrons to the positrons by high-energy photons. The method is shown to produce positron beams with 40\%-60\% polarization and low angular divergence on a femtosecond time scale. While this scheme provides a large polarization degree without necessitating high-power two-color laser pulses, a well-polarized electron beam needs to be generated in a first step which is a challenge in itself (cf. previous subsection).

\begin{figure}
    \centering
    \includegraphics[width=\textwidth]{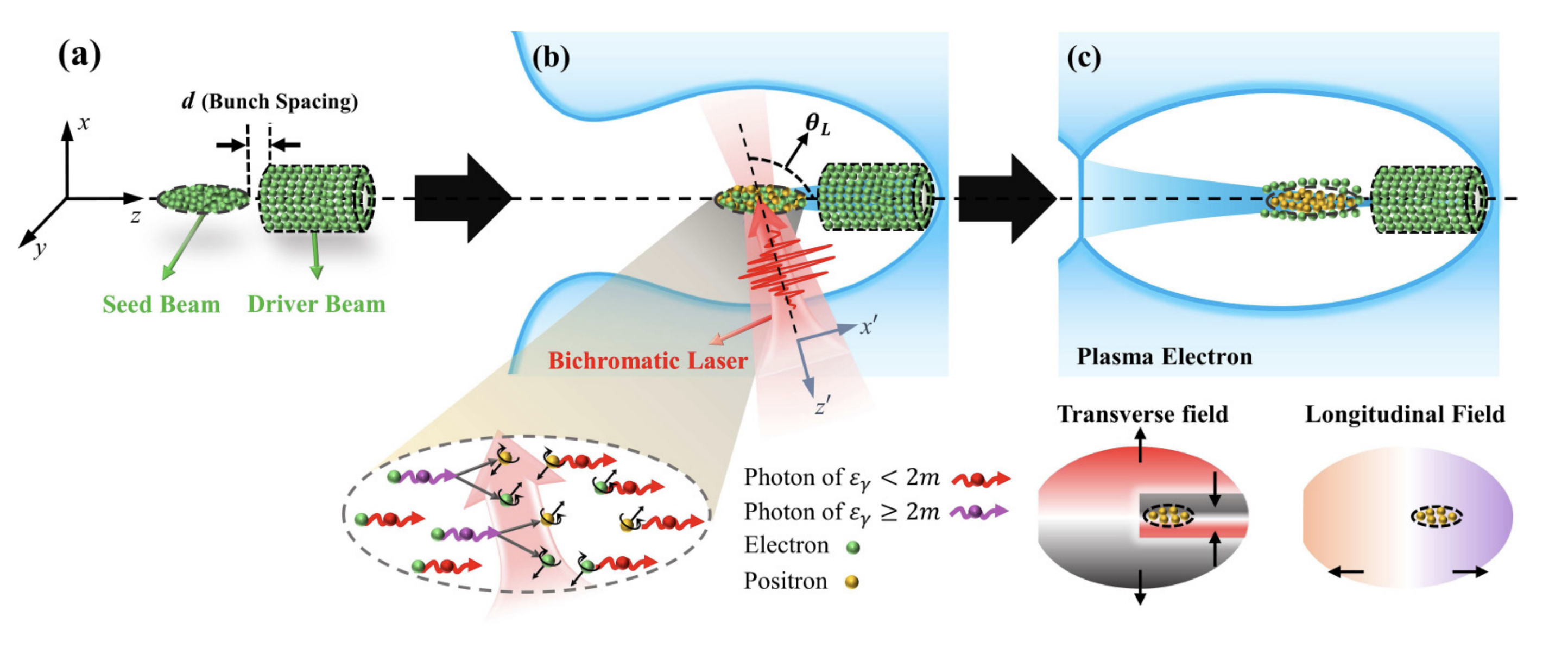}
    \caption{\label{fig:liu_gamma_splitting} (a) The hollow driver and an electron seed beam propagate in $z$-direction. (b) The driver beam generates the wakefield structure, and the seed is irradiated by a two-color laser pulse in order to produce positrons. (c) The polarized positrons are then trapped and accelerated in the wakefield. Reproduced under the terms of the CC-BY license from \cite{Liu2022_photon_splitting}. Copyright 2022, The Authors, published by American Physical Society.}
\end{figure}

Liu \textit{et al.} found the polarization of intermediate gamma quanta to be of significance for the polarization of electron-positron pairs produced in the non-linear Breit-Wheeler process \cite{Liu2022_photon_splitting}. The setup is shown in Fig. \ref{fig:liu_gamma_splitting}. They find that an average polarization of 30\% can be obtained, and show that the resulting beam can be accelerated in a wakefield driven by a hollow electron beam without significant depolarization. 

A method utilizing laser-solid interaction has been proposed by Xue \textit{et al.} \cite{Xue2023_positron_foil}. A high-intensity laser pulse is used to ionize a solid foil. The accelerated electrons are accelerated and generate gamma quanta via non-linear Compton scattering which decay into polarized positrons in a quasi-static magnetic field. The angle of irradiation is found to determine the configuration of the quasi-static field, which in turn, allows tuning of the polarization degree. The scheme delivers polarization on a 70\%-level with 0.1 nC per shot. 
Such solid-target based approaches are simpler to realize as there are no requirements for pre-polarized electron bunches, however, high-intensity laser pulses in the range of $10^{23} \;\mathrm{W/cm}^2$ are required, pushing the realization of this setup into near future.
Zhao \textit{et al.} investigated cascaded Compton scattering and Breit-Wheeler pair production on a more general level with respect to polarization transfer in \cite{Zhao2023_cascade_polarized}.

Another approach to generating polarized positrons was presented by Zhu \textit{et al.} in \cite{Zhu2024_positrons_solid}, using a similar setup to their work on electrons \cite{Zhu2024_electrons_target}: a relativistic electron beam is used to irradiate a solid target surface, inducing strong, asymmetric magnetic fields. This leads to radiative spin flips and abundant multi-photon processes, generating polarized positrons. In simulations, they find that a multi-GeV positron beam with nC charge and approx. 40\% can be produced following this approach. This setup does not require a pre-polarized beam, or high-intensity laser pulses, however, it requires a high-energy electron beam (20 GeV in the case of the presented simulations) that needs to be focused to high density.

The efficient acceleration of positrons using plasma-based methods is -- even without considering spin polarization -- challenging \cite{Cao2024_positrons_review}, and still requires significant research efforts for applications like electron-positron colliders \cite{Gessner2025_10TeV}.
A study by Dou \textit{et al.} proposed that polarized positrons can be efficiently captured and accelerated from a polarized pair plasma using multilayer microhole-array films \cite{Dou2025_positrons_microhole}. This method preserves polarization on a 90\%-level while delivering strong accelerating gradients in the range of TeV/m. While the polarization and energy levels are promising, this method relies on producing a high-quality electron-positron pair plasma first.

\section{Polarized ions} \label{sec:ions}

While single-step methods of polarization and acceleration have been proposed for polarized electron beams, no such options are currently known for ion beams. This means in particular that any theoretical research on ions has considered maintaining a high degree of spin polarization from initially polarized targets. In this section, we describe the progress on polarized ion beams from laser-plasma interaction.

A first experimental study by Raab \textit{et al.} showed that no net polarization is obtained during laser-solid interaction \cite{Raab2014_pol_measurement}.
The first proof-of-principle experiment on laser-based acceleration of polarized ${}^3$He was realized by Zheng \textit{et al.} at the PHELIX facility in GSI Darmstadt \cite{Zheng2023_evidence_he3}. The experiment was conducted with a 50 J laser pulse and a duration of 2.2 ps. The laser parameters were found to be optimal in a preceding simulation study by Engin \textit{et al.} for the helium target \cite{Engin2019_he3}. The ${}^3$He target has a density of $10^{19}$ cm$^{-3}$ and is extensively described in \cite{Fedorets2022_he3}. Due to the density and length of the target, the ions were mostly expelled in the direction transverse to laser propagation and reached energies of several MeV.
Further simulations by Gibbon \textit{et al.} indicated that shortening the interaction length could increase the amount of ions being accelerated in forward direction \cite{Gibbon2022_he3}.

In the following, several theoretical results on ion acceleration for both current and near-future laser parameters are discussed. As the preparation of solid-state based, pre-polarized targets is difficult, the theoretical results mostly consider targets with near-critical density.

\subsection{Magnetic Vortex Acceleration}

Due to the restrictive target parameters currently available for polarized ion sources, Magnetic Vortex Acceleration (MVA) is one of the few feasible acceleration mechanisms.
The process was first described by Bulanov \textit{et al.} in \cite{Bulanov2005_mva} without considering spin polarization. A more detailed analysis of the MVA process can be found in the work by Park \textit{et al.} \cite{Park2019_mva}.

In the context of polarized protons, MVA was first investigated by Jin \textit{et al.} in \cite{Jin2020_mva} by means of particle-in-cell simulations. They investigated the interaction of a multi-PW laser pulse in the range $a_0 = 25 - 100$ with a pre-polarized HCl target. The target has an electron density of $0.36 n_\mathrm{crit}$ and has a flat-top profile with steep edges.
In the case of $a_0 = 25$, 0.26 nC of protons with up to 53 MeV and a polarization of 82\% were observed. An increase in laser energy was shown to reduce the polarization: for a laser pulse with $a_0 = 100$, 3.1 nC were accelerated to 152 MeV, however with a lower polarization of only 56\%.

A publication by Reichwein \textit{et al.} studied the effect of density-down ramps at the end of the target on the final beam quality \cite{Reichwein2021_downramp}. The same HCl target was used, however with different lengths of down-ramps in the range of 0-80 {$\mu$}m. The polarization of the final proton beam was shown to remain largely robust against changes of the ramp length. Longer ramps mostly affected the spatial focusing of the beam as the transverse extent of the plasma channel and, in turn, the induced fields are changed. Ramps of significant length can, however, lead to reduced beam polarization, since a longer ramp changes the focusing of protons into the beam, introducing also particles of different spin orientation.
Zou \textit{et al.} showed in a further publication that the MVA process can be aided by a laser-driven plasma bubble for petawatt-level laser pulses \cite{Zou2025_enhanced_MVA}. In simulations, they find that monoenergetic proton beams with hundreds of MeV, hundreds of pC charge and polarization on the level of tens of percent can be produced.

Since the ions in MVA interact with the strong laser and plasma channel fields directly, the degree of beam polarization is strongly limited by the intensity of the laser pulse. In agreement with the theoretical scaling laws of Thomas \textit{et al.} \cite{Thomas2020_scaling}, higher intensities lead to higher beam energy but at the cost of reduced polarization.
Thus, in \cite{Reichwein2022_dual_pulse} a dual-pulse MVA setup was introduced to mitigate these effects. Two laser pulses with a carrier envelope phase difference of $\pi$ propagate through the target side-by-side with some transverse displacement. Both of the pulses induce their own MVA process, i.e. two plasma channels will be formed.

\begin{figure}
\centering
\includegraphics[width=0.65\textwidth]{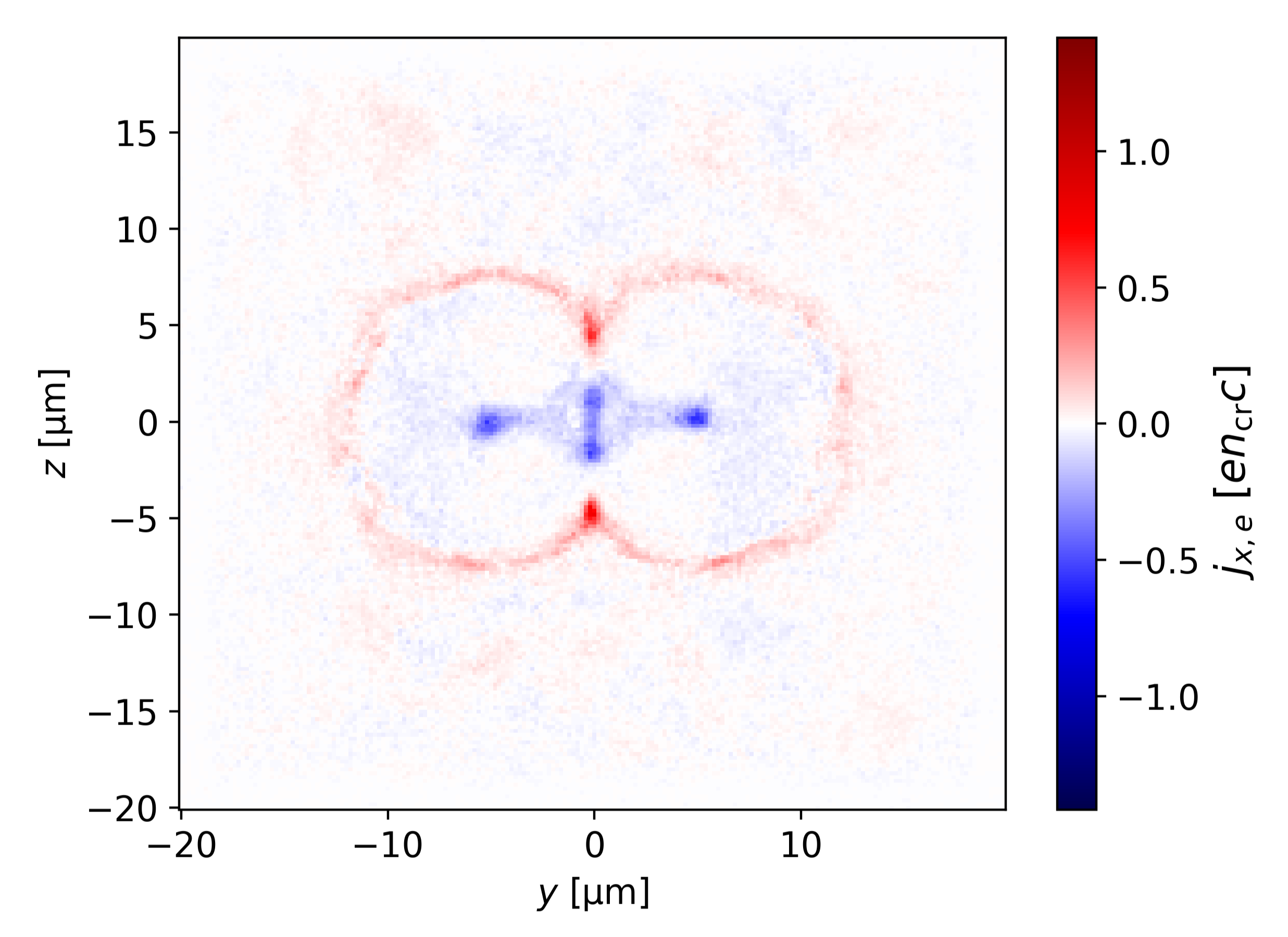}
\caption{\label{fig:reichwein2022_mva_jx} Transverse cut of the current density profile of the dual-pulse MVA scheme. Note the two distinct plasma channels with an accelerating region in their middle. Reproduced under the terms of the CC-BY license from \cite{Reichwein2022_dual_pulse}. Copyright 2022, The Authors, published by American Physical Society.}
\end{figure}

In the space between the two plasma channels, an accelerating region with weaker depolarizing fields is formed (cmp. current density in Fig. \ref{fig:reichwein2022_mva_jx}). This leads to the formation of a central filament which obtains energies similar to the comparable single-pulse MVA setup, but with higher polarization.
In \cite{Reichwein2022_dual_pulse}, polarization of 77\% was obtained for two pulses with $a_0 = 100$, while a single-pulse setup of the same total laser energy only led to 64\%. The dual-pulse scheme was also shown to improve on the angular spectrum of the accelerated beam, as instabilities were mitigated by the presence of the second laser pulse.
While the dual-pulse scheme improves on polarization compared to conventional MVA, its experimental realization is more complicated due to the necessity to align to high-power laser pulses sufficiently.

\subsection{Collisionless Shock Acceleration}

In MVA, the polarization of the ion beam decreases since the oscillating laser fields directly interacting with the ions. This is different in Collisionless Shock Acceleration (CSA), where a more homogeneous field is used to accelerate the ions. In the context of polarized ion beams, multiple setups utilizing a solid foil in front of the near-critical density polarized target were proposed. When the laser pulse irradiates the solid foil, it heats up the electrons and induces a shock wave with rather homogeneous electric field. The gaseous component can then be reflected by the shock wave and accelerated to high energies. The general mechanism of CSA is detailed, e.g., in \cite{Silva2004_shock}.

In \cite{Yan2022_csa_foil}, Yan and Ji first suggested the use of CSA for polarized proton beams. Their setup consisted of a laser pulse of $a_0 = 20-80$ irradiating a 2 {$\mu$}m Carbon foil which is placed in front of a near-critical, polarized HCl target. Proton beams with tens of MeV and a polarization on a 90\%-level were obtained.
A subsequent study by Yan \textit{et al.} investigated the use of a microstructured foil \cite{Yan2023_csa_microstructure}, where the structures were utilized to increase the temperature and the number of hot electrons.  Accordingly, the energy of the reflected protons was increased.

\begin{figure}
    \centering
    \includegraphics[width=0.65\linewidth]{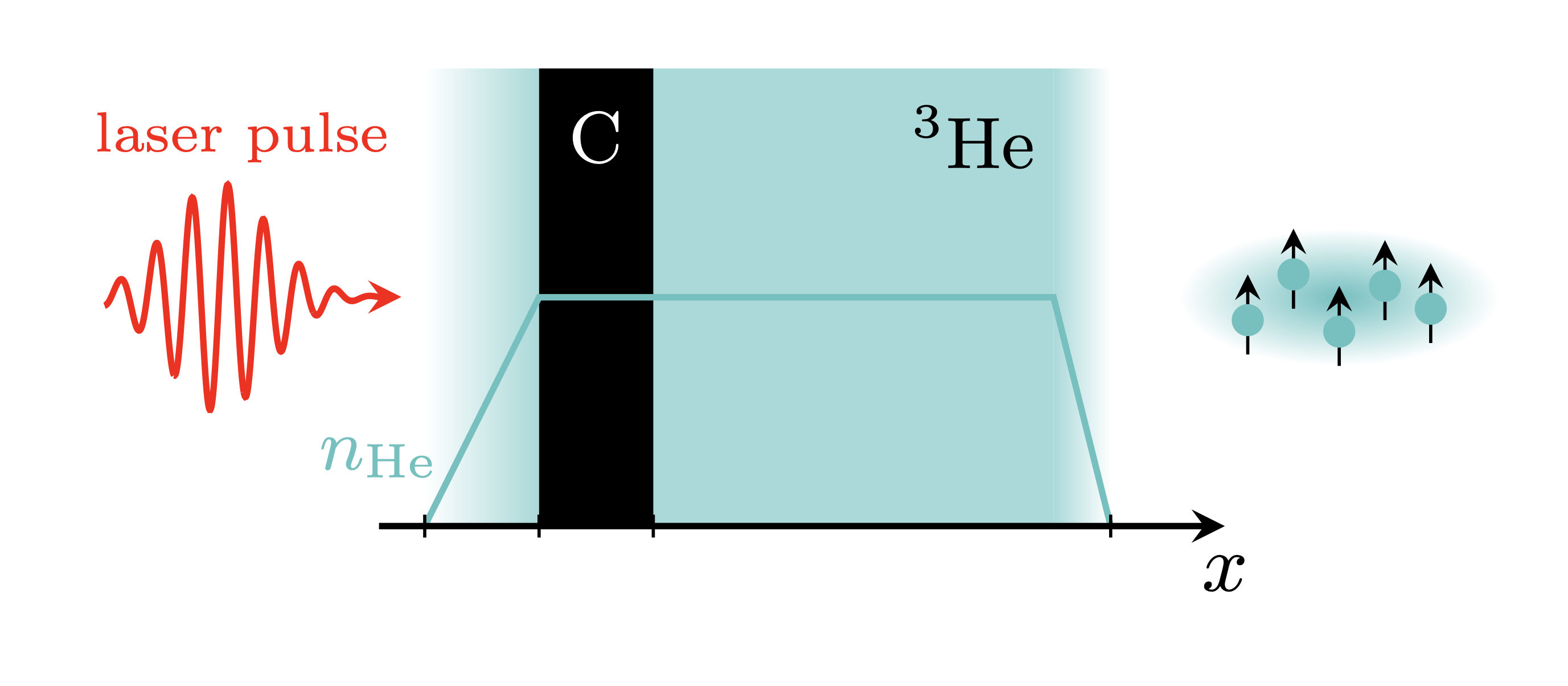}
    \caption{\label{fig:reichwein2024_csa} Schematic of the setup for Collisionless Shock Acceleration. A solid Carbon foil is placed in front of the near-critical density, polarized Helium-3 target. Reproduced under the terms of the CC-BY license from \cite{Reichwein2024_csa}. Copyright 2024, The Authors, published by IOP Publishing Ltd.}
\end{figure}

The aforementioned studies have been extended by Reichwein \textit{et al.} to the case of polarized ${}^3$He and to higher laser intensities \cite{Reichwein2024_csa}. A schematic of the setup is shown in Fig. \ref{fig:reichwein2024_csa}. Similarly to the results with protons, high polarization is achieved with CSA, in particular, it remains in the range of 90\% for intensities of $a_0 = $ 100-200. Moreover, the publication investigated the influence of radiation reaction on the acceleration process. As the electrons radiate part of their energy, the electromagnetic field structure of the system and, subsequently, the ion motion is changed. This leads to a reduction in energy, but an increase in beam charge, since more Helium ions can be trapped inside the accelerating field. Still, the beam charge remains in the range of tens of pC, significantly lower than the (up to) few nC that MVA can provide. The polarization remains largely unaffected by radiation reaction.

An alternative pathway to CSA was suggested by Yan \textit{et al.} in \cite{Yan2022_csa_co2}, where no additional foil is necessary to drive the process. A CO$_2$ laser wavelength of 10 {$\mu$}m was used, which gives a critical plasma density comparable to the density achieved in pre-polarized targets created by photo-dissociation. This allows a laser pulse in the intensity range of $10^{17}-10^{18}$ W/cm$^2$ to be used. The shock front then is created by the long laser pulse (3 ps) that piles up protons and a spectral peak energy of approximately 2 MeV was found, with a cut-off energy of about 7 MeV. The polarization remained $> 80\%$ in a large parameter range.

\subsection{Other mechanisms}

Hützen \textit{et al.} also considered the acceleration of ions using a bubble-channel structure for multi-PW laser systems \cite{Huetzen2020_bubble_ion}. The acceleration scheme is based on work by Shen \textit{et al.} \cite{Shen2007_bubble_ion}. Both HCl and HT plasma with densities in the range of $10^{19}$-$10^{21}$ cm$^{-3}$ and a length of 600 $\mu$m are considered. The plasma is irradiated by a high-intensity laser pulse of up to $a_0 = 223$, leading to the formation of a wide plasma channel. At the leading edge of this channel structure, polarized particles are piled up and accelerated to energies in the multi-GeV range. The investigation of this mechanism was extended by Li \textit{et al.} in \cite{Li2021_proton_bubble}.
Here the protons are pre-accelerated by a CP laser pulse, and subsequently trapped and further accelerated in the front region of a wakefield bubble. The ratio of hydrogen and tritium is found to be crucial for the efficiency of the acceleration. If chosen correctly, a polarized proton beam can be accelerated up to GeV-level in the wakefield structure. 
The proposed setup is, however, currently beyond experimental capabilities regarding the required density regime as well as the laser intensity.

\begin{figure}
    \centering
    \includegraphics[width=\linewidth]{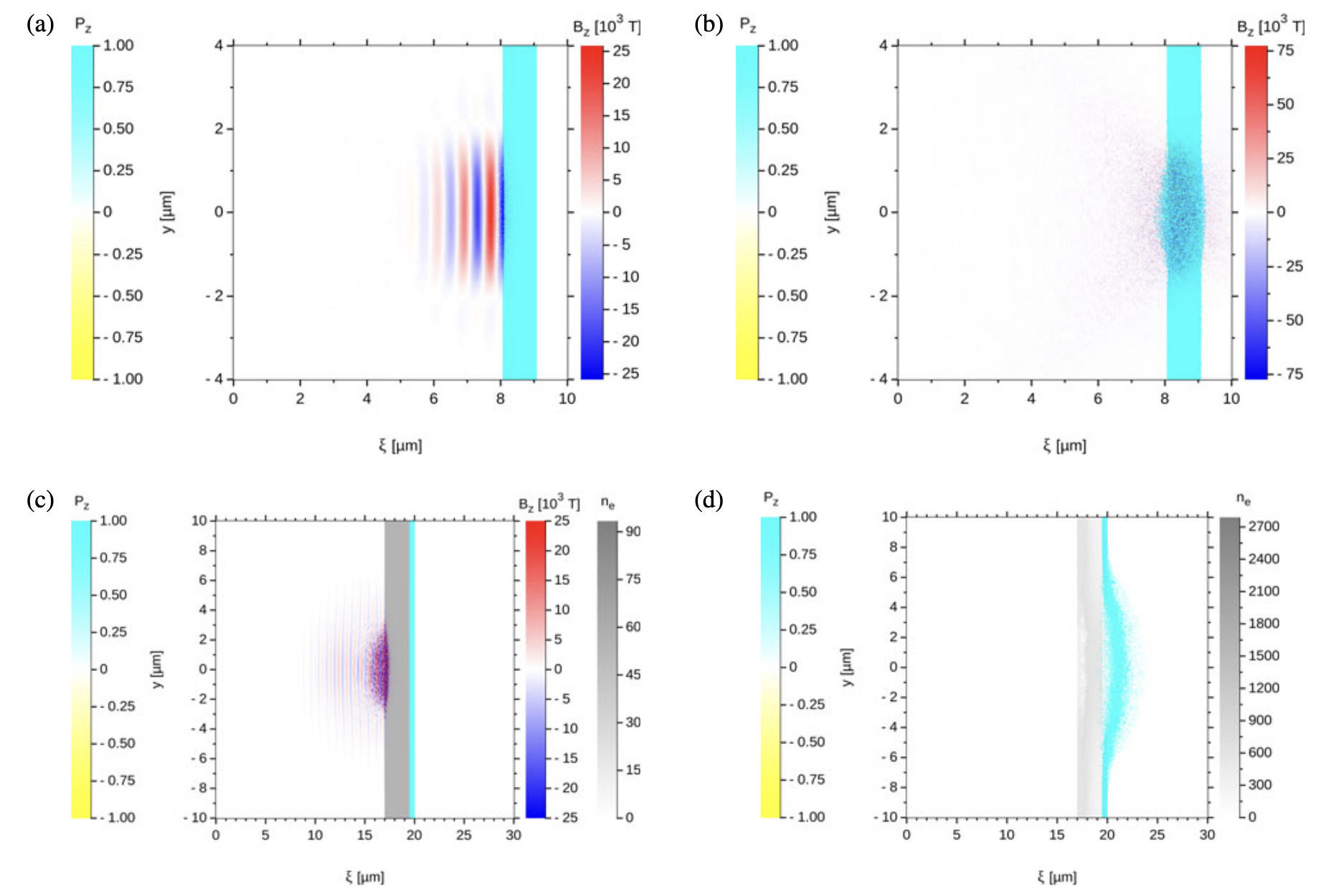}
    \caption{\label{fig:huetzen2019_foils}Proton spin polarization for the pure Hydrogen foil (a)-(b), as well as the Aluminium-Hydrogen composite target (c)-(d). Reproduced under the terms of the CC-BY license from \cite{Huetzen2019_protons}. Copyright 2019, The Authors, published by Cambridge University Press.}
\end{figure}

The acceleration of polarized ions from hypothetical, overdense targets was considered by Hützen \textit{et al.} in \cite{Huetzen2019_protons}. Two sets of PIC simulations were carried out: at first a fully polarized hydrogen layer of 1 {$\mu$}m thickness and $128 n_\mathrm{crit}$ density was considered (see Fig. \ref{fig:huetzen2019_foils}). The second set of simulations considered a 2.5 {$\mu$}m aluminium layer at $35 n_\mathrm{crit}$ with a shorter polarized proton layer with 0.5 {$\mu$}m thickness. The dominant acceleration mechanism in both setups was found to be Target Normal Sheath Acceleration (TNSA). In the former case, preservation of proton polarization for up to 0.24 ps was observed, while for the latter case it was found to be preserved for at least 0.18 ps.
While acceleration mechanisms like TNSA or radiation pressure acceleration (RPA) would generally be of interest due to their beam quality, they are currently ruled out as preparing pre-polarized, solid targets is unfeasible, leaving MVA and CSA as the main candidates.

\section{Polarized gamma quanta} \label{sec:gamma}

In this section, we discuss the generation of polarized gamma-ray photons resulting from the interaction of high-intensity laser pulses with plasmas or charged-particle beams. 
The study of such polarized photons is motivated both by their wide-ranging practical applications and by their potential role in advancing multi-messenger studies of laser-driven plasmas.
Polarized gamma-ray photons have practical applications that include nuclear physics~\cite{moortgat2008polarized}, high energy physics~\cite{speth1981giant,uggerhoj2005interaction}, and astrophysics~\cite{laurent2011polarized,boehm2017circular}.
Moreover, the photon polarization can provide an additional degree of freedom of information to help retrieve the in-situ electron dynamics in ultrarelativistic laser or beam driven plasmas~\cite{Gong2022_in_situ}.

\subsection{Polarized Multi-GeV gamma-photon beams via Single-Shot Laser-Electron Interaction}

\begin{figure}
\centering
\includegraphics[width=0.6\textwidth]{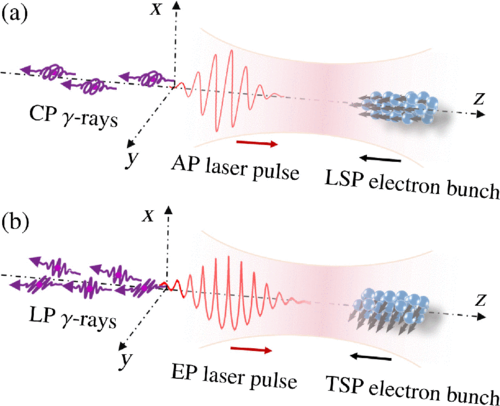}
\caption{Scenarios of generating CP and LP $\gamma$-rays via nonlinear Compton scattering. (a) An arbitrarily polarized (AP) laser pulse propagating along $+z$ direction and head-on colliding with a longitudinally spin-polarized electron bunch produces CP $\gamma$-rays rays. (b) An elliptically polarized laser pulse propagating along $+z$ direction and colliding with a transversely spin-polarized (TSP) electron bunch produces LP $\gamma$-rays. The major axis of the polarization ellipse is along the $x$ axis. Reproduced with permission from \cite{Li2020_gamma}. Copyright 2020, American Physical Society.}\label{fig:Figures_chapter5_fig_2020Li_photon}
\end{figure}

Recently, by considering the photon polarization effect, Li \textit{et al.} \cite{Li2020_gamma} investigated the generation of circularly polarized (CP) and linearly polarized (LP) $\gamma$-rays via the interaction of an ultraintense laser pulse with a spin-polarized counter-propagating ultrarelativistic electron beam. Pre-polarized energetic electrons collide with the laser pulse to experience nonlinear Compton scattering in the quantum radiation dominated regime as shown in Fig.~\ref{fig:Figures_chapter5_fig_2020Li_photon}.
The study develops a Monte Carlo method based on the locally constant field approximation (LCFA) that employs electron-spin-resolved probabilities for polarized photon emissions.
Li \textit{et al.} \cite{Li2020_gamma} show efficient ways for the transfer of electron polarization to high-energy photon polarization. The results demonstrate that multi-GeV CP (LP) $\gamma$-rays with polarization of up to about 95\% can be generated by a longitudinally (transversely) spin-polarized electron beam, with a photon flux meeting the requirements of recent proposals for the vacuum birefringence measurement in ultra-strong laser fields. The high-energy, high-brilliance, and high-polarization $\gamma$-rays generated in this process are also beneficial for other applications in high-energy particle physics, light sources, and laboratory astrophysics. Recently, using the quantum operator method by Baier and Katkov to calculate the polarization-resolved probabilities within the quasi-classical approach and LCFA, Chen \textit{et al.} \cite{chen2022electron} presented a detailed derivation of fully polarization-resolved probabilities for high-energy photon emission and $e^-e^+$ pair production in ultra-strong laser fields. These probabilities, accounting for both electron spin and photon polarization of incoming and outgoing particles, are crucial for the development of QED Monte Carlo simulations and QED-particle-in-cell codes. 

In subsequent studies, Wan \textit{et al.} \cite{wan2020high} investigated the interaction between an unpolarized 10\,GeV electron beam and a counter-propagating linearly polarized laser pulse (with a peak intensity $a_0\approx50$ and pulse duration $\tau\approx 50\,$fs) using the same semi-classical Monte Carlo approach. They found that high-energy, linearly polarized $\gamma$-ray photons are generated during the interaction through nonlinear Compton scattering, with an average polarization degree exceeding 50\%. These photons further interact with the laser fields, leading to electron-positron pair production via the nonlinear Breit-Wheeler process. Photon polarization is found to enhance the pair production yield by more than 10\%. 
To understand the influence of photon polarization on pair production, Dai \textit{et al.} \cite{dai2022photon} used fully polarization-resolved Monte Carlo simulations to examine the correlation between photon and electron (positron) polarization in the multiphoton Breit-Wheeler process, where a 10\,GeV electron beam is set to collide with a counter-propagating elliptically polarized laser pulse with a peak intensity $a_0=100$, pulse duration $\tau\approx 27\,$fs, and the ellipticity $\epsilon=0.03$. They found that the polarization of the produced $e^-e^+$ is reduced by 35\% and the pair yield is decreased by 13\% if the intermediate photon polarization is taken into account.

In addition, Lv \textit{et al.} proposed to generate brilliant polarized $\gamma$-photon beams via the vacuum dichroism-assisted vacuum birefringence effect, using an unpolarized electron beam \cite{Lv2025_gamma_birefringence}. They consider that a linearly polarized laser pulse is split into two sub-pulses, with the first one colliding with a dense unpolarized electron beam to generate a linearly polarized $\gamma$-photon beam, which then further collides with the second sub-pulse and is transformed into a circularly polarized one via the vacuum birefringence effect.

\subsection{Polarized high-energy brilliant gamma-ray sources via laser-plasma interaction}
\begin{figure}
\centering
\includegraphics[width=0.6\textwidth]{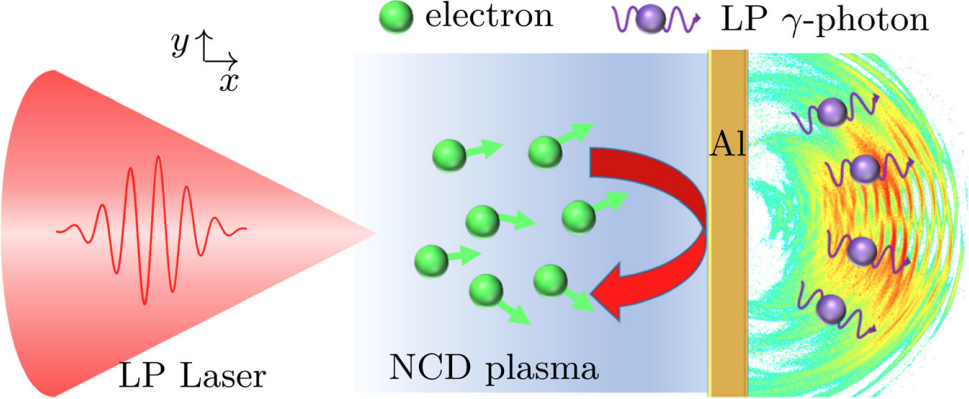}
\caption{Scenario for the generation of LP $\gamma$-ray photons via nonlinear Compton scattering. An ultra-strong LP laser pulse, polarized along the $y$-axis and propagating along the $x$-direction, irradiates a near-critical density hydrogen plasma followed by an ultra-thin Al foil. Reproduced under the terms of the CC-BY license from \cite{Xue2020_generation}. Copyright 2020, The Authors, published by AIP Publishing.}\label{fig:Figures_chapter5_fig_2020Xue_MRE}
\end{figure}

It is worth emphasizing that the generation of polarized $\gamma$-ray photons via single-shot laser-electron or beam-beam interaction has been studied based on a test-particle simulation~\cite{Li2020_gamma, chen2022electron, wan2020high, dai2022photon, Gao_2024PhysRevA.110.013502}. This means that collective effects of charged particles, such as self-generated plasma electric and magnetic fields, are not taken into account. 
Recently, Xue \textit{et al.} \cite{Xue2020_generation} studied the generation of highly-polarized high-energy brilliant $\gamma$-rays via laser-plasma interaction, where plasma collective effects are self-consistently considered by implementing the electron-spin-resolved probabilities for polarized photon emission in PIC simulations. 
By utilizing PIC simulations, Xue \textit{et al.} \cite{Xue2020_generation} show that when an ultraintense LP laser pulse irradiates a near-critical-density (NCD) plasma target followed by an ultra-thin aluminum (Al) foil, the electrons in the NCD plasma are first accelerated by the driving laser via direct laser acceleration to ultrarelativistic energies and then collide with the light pulse reflected by the Al foil. 
The PIC simulations of this all-optical scheme show that brilliant LP $\gamma$-ray photons are produced via nonlinear Compton backscattering with photon energy up to hundreds of MeV and an average LP degree of $\sim70\%$. This high LP degree is reasonable, since the driving laser pulse is a fully polarized light with a LP degree of $100\%$.
Alternatively, instead of using the double-layer target in Fig.~\ref{fig:Figures_chapter5_fig_2020Xue_MRE} to generate $\gamma$-ray photons via nonlinear Compton backscattering, Xue \textit{et al.} \cite{Xue2020_generation} also calculate photon emission by considering a conical gold target filled with an NCD hydrogen plasma. When ultraintense LP light impinges on the NCD target inside the Au cone, the bulk plasma electrons are pushed forward and the consequent current density sustains a strong quasi-static self-generated magnetic field. 
For forward moving electrons, the self-generated magnetic field provides a restoring force to enable electrons' betatron oscillation inside a guided plasma channel in the conical Au target, in which emitted $\gamma$-ray photons concentrate in the angular region $\theta< 20^\circ$, and the cutoff energy exceeds 450 MeV. In this scheme, the LP degree can reach $\sim 50\%$ for the energetic part of emitted photons.

\begin{figure}
\centering
\includegraphics[width=0.6\textwidth]{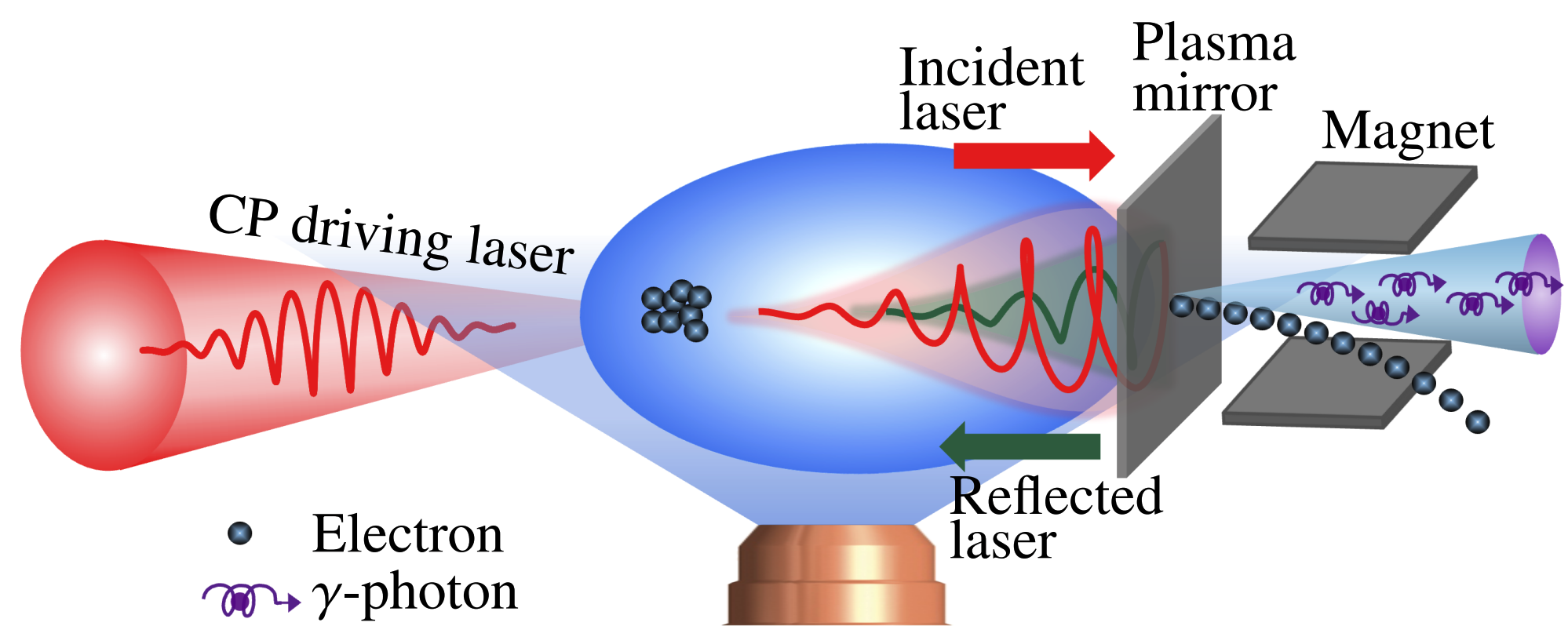}
\caption{Schematic for all-optical generation of brilliant CP
$\gamma$-ray beam via single-shot laser plasma interaction by combining a plasma mirror and an electron beam produced from wakefield acceleration. Reprinted with permission from \cite{Wang_2022brilliant}. Copyright 2022, Optica Publishing Group.}\label{fig:Figures_chapter5_fig_2022Wang_OL}
\end{figure}

By employing a similar scheme of the combination of laser wakefield acceleration and plasma mirror as shown in Fig.~\ref{fig:Figures_chapter5_fig_2022Wang_OL}, Wang \textit{et al.} \cite{Wang_2022brilliant} found that a brilliant CP $\gamma$-ray beam could be generated in a weakly nonlinear Compton scattering regime with moderate laser intensities of $a_0\approx 3$, where the helicity of the driving laser pulse is transferred to the emitted $\gamma$-photons. The authors claim that the calculated average polarization degree of $\gamma$-photons reaches $\sim61\% (20\%)$ with a peak brilliance of $\gtrsim 10^{21}$ photons/(s$\cdot$ mm$^2$$\cdot$ mrad$^2$$\cdot$ 0.1\% BW) around 1 MeV (100 MeV) \cite{Wang_2022brilliant}. 
Recently, Wang \textit{et al.} also investigated the mechanism of transfer of spin angular momentum during the transition from linear to nonlinear processes \cite{wang2024manipulation}. Their findings suggest that to achieve high-energy, high-brilliance, and high-polarization gamma rays, increasing the laser intensity is crucial for an initially spin-polarized electron beam. However, for an initially unpolarized electron beam, it is also necessary to increase the electron beam's energy, in addition to boosting laser intensity.
In addition, Qian \textit{et al.} \cite{qian2025bright} reported a numerical study that a flash of x/$\gamma$-ray photons with energy $>10\,$keV and $65\%$ degree of LP could be produced by firing a laser of intensity $10^{21}\mathrm{W/cm}^{2}$ on a solid Al target.
Furthermore, Cui \textit{et al.} proposed a theoretical scheme for producing collimated high-brilliance polarized $\gamma$-rays with attosecond duration via the irradiation of a GeV electron beam on a solid-density plasma \cite{Cui2025_gamma_instabilities}.

\subsection{Application: deciphering in situ electron dynamics of high-intensity laser-driven plasma}

From Li \textit{et al.}~\cite{Li2020_gamma} and Xue \textit{et al.}~\cite{Xue2020_generation}, one can find that $\gamma$-ray photons with high LP degree of $P_\mathrm{LP}>70\%$ can be produced via nonlinear Compton backscattering. These highly polarized $\gamma$-ray photons have various applications in laboratory astrophysics and high-energy physics.
On the other side, the new degree of freedom of information provided by photon polarization indicates that $\gamma$-photon polarization could be a useful tool for retrieving the in situ transient dynamics of plasma electrons. 
The successful decoding of field properties near the event horizon of black holes~\cite{BH_image_2021_March} has re-stimulated the interest in measurements based on photon polarization~\cite{lembo2021cosmic}. Although polarized light is vulnerable to magneto-optic disturbance~\cite{faraday1846magnetization}, the high-frequency $\gamma$-photon is robust during penetration of the plasma depth \cite{chen2012introduction}. Previously, observation of celestial $\gamma$-ray emission has helped to understand star-forming galaxies~\cite{roth2021diffuse}, accretion flows around black holes~\cite{kimura2021soft}, and active galactic nuclei~\cite{murase2020hidden}. In contrast to routinely detected quantities of arrival time, direction, and energy, the $\gamma$-photon polarization (GPP), provides new insights on the relativistic jet geometry~\cite{zhang2019detailed} and magnetic field configuration~\cite{gill2020linear}, which allows identification of cosmic neutrino scattering~\cite{batebi2016generation}, dark matter annihilation~\cite{boehm2017circular}, and acceleration mechanisms surrounding crab pulsars~\cite{dean2008polarized}. 
All this progress appeals to a laboratory platform for simulating and examining the GPP associated with the plasma phenomena.

\begin{figure}
\centering
\includegraphics[width=0.9\textwidth]{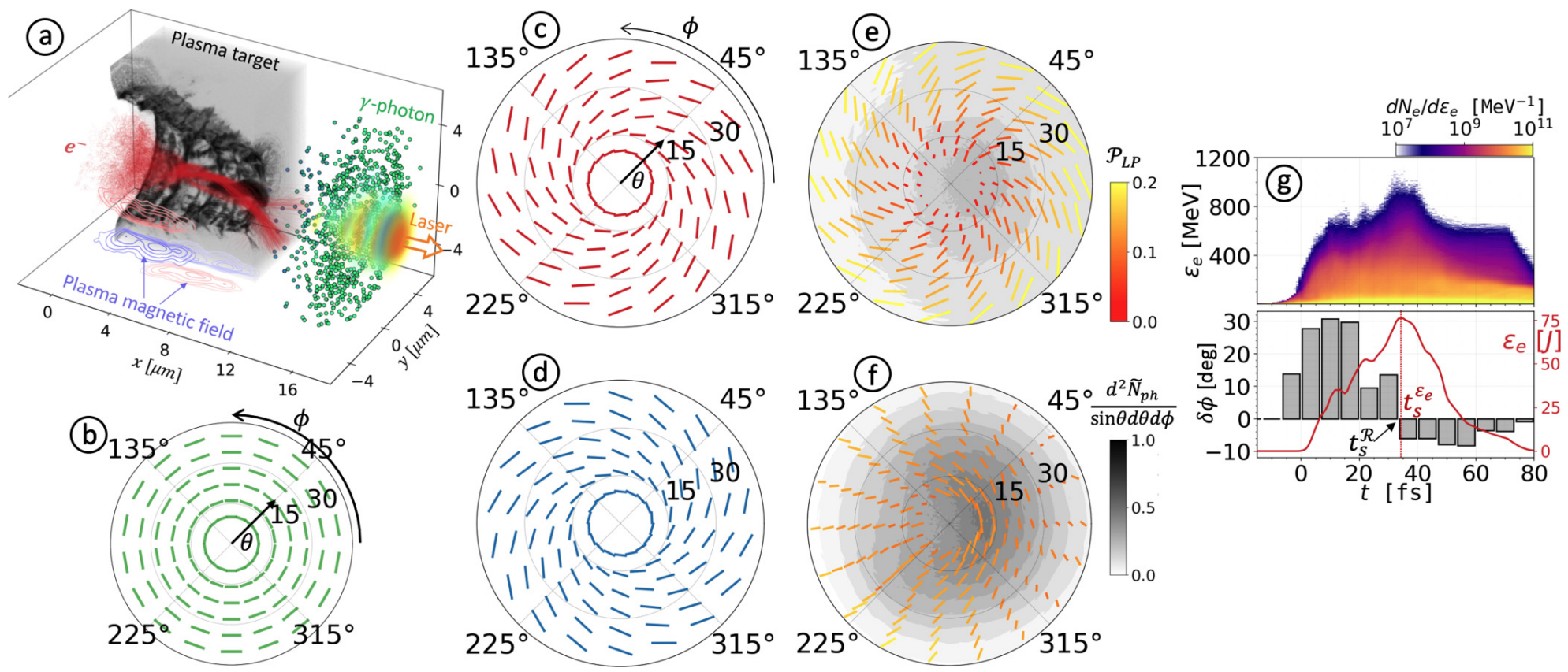}
\caption{The schematic for $\gamma$-ray photon emissions from a plasma interacting with a laser pulse. (b) The LP
orientation is along the local azimuthal direction. The analytically predicted LP orientation for the electron undergoing acceleration (c) and deceleration (d). The simulated $\gamma$-photon LP for the time at (e) $t = 10$ and (f) 40 fs, with the LP degree $P_\mathrm{LP}$ and the normalized number distribution $d^2N_{ph}/\sin\theta d\theta d\phi$. 
(g) The time evolution of the electron energy spectrum $dN_e/d\varepsilon_e$, the polarization orientation $\delta\phi$, and the kinetic energy $\varepsilon_e$. Reproduced under the terms of the CC-BY license from \cite{Gong2022_in_situ}. Copyright 2022, The Authors, published by American Physical Society.}\label{fig:Figures_chapter5_fig_2022Gong_PRR}
\end{figure}

Recently, Gong \textit{et al.}~\cite{Gong2022_in_situ} found that the angular pattern of $\gamma$-photon linear polarization is explicitly correlated with the dynamics of the radiating electrons, which provides information on the laser-plasma interaction regime.
Using 3D-PIC simulations, Gong \textit{et al.}~\cite{Gong2022_in_situ} studied the polarization-resolved $\gamma$-photon emission in plasma driven by a circularly polarized laser pulse [see Fig.~\ref{fig:Figures_chapter5_fig_2022Gong_PRR}]. The simulation results showed that the collective orientation of the $\gamma$-photons' LP resembles a spiral shape with the rotation tendency determined by the acceleration status of the radiating electrons. To characterize the degree of rotation tendency, Gong \textit{et al.} introduced the spiral tendency $\delta\phi$, as the deviation of the orientation of $\gamma$-photon LP with respect to the azimuthal direction. 

%\textcolor{red}{DEFINE OTHER PARAMETERS}
In the moderate QED regime $\chi_e<1$, the direction of the emitted $\gamma$-photon LP is primarily parallel to the acceleration direction perpendicular to the electron momentum, $\textbf{a}_\perp\equiv\textbf{a}-(\textbf{a}\cdot\hat{\textbf{v}}) \hat{\textbf{v}}$, where the hat symbol denotes the unit vector. Thus, the polarization orientation is derived as $\hat{a}_{\perp,y} \approx -\sin\phi\left(\frac{\Gamma}{\epsilon\gamma_e}+\frac{\kappa_b\cos\theta}{\epsilon\Gamma}\right) -\frac{(-\bm{\beta}\cdot\textbf{E})\cos\phi}{\gamma_e}$ and $\hat{a}_{\perp,z} \approx \cos\phi\left(\frac{\epsilon\Gamma}{\gamma_e}+\frac{\epsilon\kappa_b\cos\theta}{\Gamma}\right)-\frac{(-\bm{\beta}\cdot\textbf{E})\sin\phi}{\gamma_e}$, where $\bm{\beta}=\textbf{v}/c$, $\theta=\arctan[(p_y^2+p_z^2)^{1/2}/p_x]$ and $\phi=\mathrm{arctan2}(p_z,p_y)$. 
Here, $\chi_e = (e \hbar / m_e^3 c^4) |F_{\mu\nu}p^\nu|$ denotes the invariant quantum parameter of the electron with the field tensor $F_{\mu\nu}$ and the electron four-momentum $p^\nu$ and $\Gamma \equiv \gamma_e - (p_x / m_e v_\mathrm{vph})$ is the dephasing value of the electron within the laser field. $\kappa_b$ is the coefficient for characterizing the self-generated magnetic field and $\epsilon$ is the ellipticity of the laser electromagnetic wave. $p_{x,y,z}$ denotes the electron momentum.

When the electron energy gain is negligible, that is $-\bm{\beta}\cdot\textbf{E}=0$, the orientation of the $\gamma$-photon LP is along the azimuthal direction $\hat{\textbf{a}}_a=(-\sin\phi,\cos\phi)^\intercal$, which collectively resembles multiple concentric rings with each polarization segment along the azimuthal direction [see Fig.~\ref{fig:Figures_chapter5_fig_2022Gong_PRR}(b)].
The deviation of the $\gamma$-photon LP orientation from the azimuthal direction is quantified by $\delta\phi\in[-90^\circ,90^\circ]$, which is the relative angle between $\hat{\textbf{a}}_\perp$ and $\hat{\textbf{a}}_a$ and is calculated as
\begin{eqnarray}\label{eq:R}
\delta\phi\approx\arcsin\left\{ \frac{-\bm{\beta}\cdot\textbf{E}}{\sqrt{\left[\Gamma+(\gamma_e\kappa_b\cos\theta/\Gamma)\right]^2+(\bm{\beta}\cdot\textbf{E})^2}}\right\}.
\end{eqnarray}
If the radiating electron is undergoing acceleration with $-\bm{\beta}\cdot\textbf{E}>0$ ($-\bm{\beta}\cdot\textbf{E}<0$), the GPP orientation $\delta\phi>0$ ($\delta\phi<0$) corresponds to the counter-clockwise (clockwise) spiral tendency in the angular distribution of the $\gamma$-photon LP as shown in Fig.~\ref{fig:Figures_chapter5_fig_2022Gong_PRR}(c) [Fig.~\ref{fig:Figures_chapter5_fig_2022Gong_PRR}(d)].
The 3D-PIC simulation results of the orientation of the emitted $\gamma$-photon LP exhibit a counter-clockwise spiral tendency at $t=10$\,fs [Fig.~\ref{fig:Figures_chapter5_fig_2022Gong_PRR}(e)] reproducing well the analytical prediction for the accelerating electron. Here, the averaged polarization angle is $\delta\phi\approx 30.6^\circ$ and the LP degree is $P_\mathrm{LP}\approx 15.3\%$. The clockwise spiral tendency in Fig.~\ref{fig:Figures_chapter5_fig_2022Gong_PRR}(f) implies the deceleration of plasma electrons occurring later at $t=40\,$fs. The time-resolved $\delta\phi$ explicitly reflects that the electrons are predominantly accelerated (decelerated) at $t\lesssim t_s^{\varepsilon_e}$ ($t\gtrsim t_s^{\varepsilon_e}$) [see Fig.~\ref{fig:Figures_chapter5_fig_2022Gong_PRR}(g)], where $t_s^{\varepsilon_e}\sim 35\,$fs is the saturation time of the electron energy.
The correlation between the spiral tendency of photon LP $\delta\phi$ and electron in-situ dynamics demonstrates that the retrieving method of utilizing the information of photon polarization might be beneficial for better understanding of phenomena in broad high-intensity interaction scenarios including ion acceleration, direct laser acceleration, high-harmonic generation, brilliant photon emission, ultra-dense nanopinches, and $e^-e^+$ pair plasma cascades.

\subsection{Application: exploring electron dynamics in relativistic astrophysical plasmas}

As a versatile information carrier of multi-messenger astrophysics~\cite{bartos2017multimessenger,meszaros2019multi,komatsu2022new}, photon polarization is critical to measuring the magnetic configuration near black holes~\cite{akiyama2021first} and crab nebulae~\cite{bucciantini2023simultaneous}, and for analyzing the particle acceleration in the blazar jet~\cite{liodakis2022polarized}.
Recent studies on relativistic collisionless shocks (RCSs) have attracted great attention since RCSs could instigate electron stochastic acceleration, similar to the Fermi process~\cite{fermi1949origin,fermi1954}, which has been well recognized as sources of energetic electrons in the universe.
Therefore, the question arises of whether the polarization feature of spontaneously emitted photons can be employed to reveal some new acceleration mechanisms in an RCS.

\begin{figure}
\centering
\includegraphics[width=0.7\textwidth]{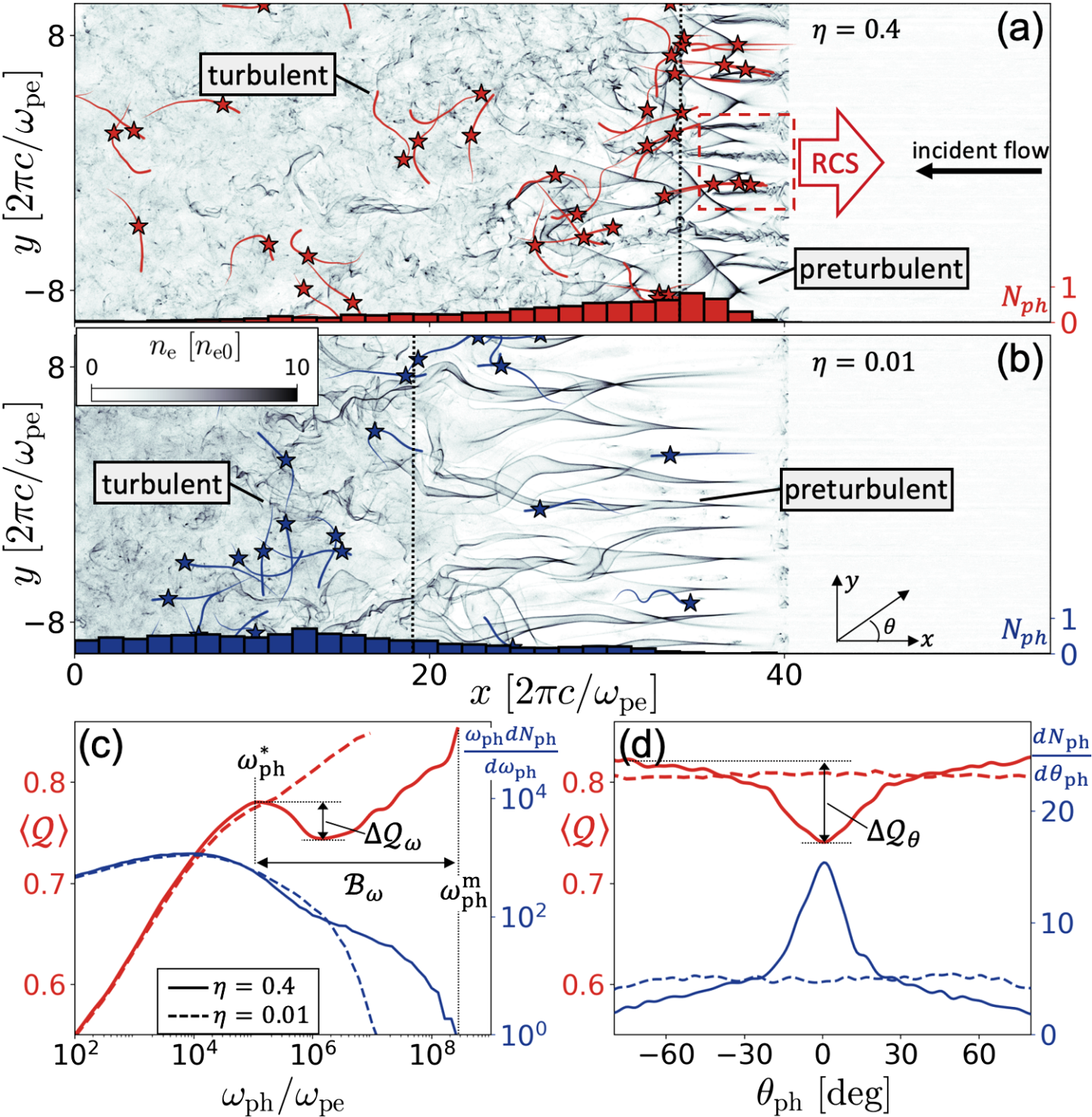}
\caption{The dynamics of a counter-streaming plasma interface: The electron density $n_e$ at $t=80\pi/\omega_{pe}$ for the case with ion fraction of (a) $\eta=0.4$ and (b) $\eta=0.01$, where lines present the typical electron moving tendency with stars marking the photon emission and the histograms display the spatial distribution of emitted photons with $\omega_{ph}>10^{-2}\omega_{ph}^m$. (c) Photon LP degree $\left<\mathcal{Q}\right>$ and energy spectra $\omega_{ph}dN_{ph}/d\omega_{ph}$ vs $\omega_{ph}$. (d) $\left<\mathcal{Q}\right>$ and $dN_{ph}/d\theta_{ph}$ vs $\theta_{ph}$. Reproduced under the terms of the CC-BY license from \cite{Gong2023_slingshot}. Copyright 2023, The Authors, published by American Physical Society.}\label{fig:Figures_chapter5_fig_2023Gong_PRL_slingshot}
\end{figure}

In a recent study, Gong \textit{et al.}~\cite{Gong2023_slingshot} found that photon polarization information can be utilized to explore the new mechanism of electron slingshot acceleration in counter-streaming plasma. 
They focused on the transient electron dynamics in the transition to turbulence near the counter-streaming interface of an unmagnetized pair-loaded relativistic collisionless shock precursor, which is potentially associated with the outflow of gamma-ray bursts (GRBs).

Gong \textit{et al.}~\cite{Gong2023_slingshot} carried out 2D simulations of counter-streaming RCSs, see Fig.~\ref{fig:Figures_chapter5_fig_2023Gong_PRL_slingshot}.
The snapshot of the electron density $n_e$ in Fig.~\ref{fig:Figures_chapter5_fig_2023Gong_PRL_slingshot}(a) shows that the filamentation exists at the front of the RCS interface. 
Between two adjacent filaments, an electron focusing point emerges, and following that, two oblique density strips stretch out [see Fig.~\ref{fig:Figures_chapter5_fig_2023Gong_PRL_slingshot}(b)]. 
Behind the strips, the coherent filaments and focusing points disappear, while the turbulence shows up. 
It should be highlighted that photons with energy $\varepsilon_{ph}\equiv \hbar\omega_{ph}> 10^{-2}\hbar\omega_{ph}^m$ are primarily emitted by electrons near the interface, where $\omega_{ph}^m\sim 10^8\omega_{pe}$ is the photon cut-off frequency and $\hbar$ is Planck's constant.
The degree of photon's linear polarization along the direction of the electron's transverse acceleration is characterized by the Stokes parameter $\mathcal{Q}$~\cite{mcmaster1954polarization}, formulated as
\begin{equation}\label{eq:Q_theory}
\mathcal{Q}= \frac{\varepsilon_e(\varepsilon_e-\varepsilon_{ph}) K_\frac{2}{3}(\zeta)}{[\varepsilon_e^2+(\varepsilon_e-\varepsilon_{ph})^2]K_\frac{2}{3}(\zeta)-\varepsilon_e(\varepsilon_e-\varepsilon_{ph})\tilde{K}_\frac{1}{3}(\zeta)},
\end{equation}
where $K_{n}(\zeta)$ is the modified secondary Bessel function, $\widetilde{K}_{1/3}(\zeta) = \int_{\zeta}^\infty K_{1/3}(z)\mathrm{d}z$, $\zeta=2\varepsilon_{ph}/[3\chi_e(\varepsilon_e-\varepsilon_{ph})]$, and $\varepsilon_e=\gamma_e m_ec^2$ the electron energy; $\chi_e\equiv(e\hbar/m_e^3c^4)|F_{\mu\nu}p^\nu|$ is the electron quantum strong-field parameter with the field tensor $F_{\mu\nu}$, and the electron four-momentum $p^\nu$. 
At $\chi_e\ll 0.1$, $\partial \mathcal{Q}/\partial \varepsilon_{ph}>0$ as predicted by Eq.~(\ref{eq:Q_theory}) manifests a monotonic dependence of $\mathcal{Q}$ on $\omega_{ph}$, because for the higher-frequency radiation the formation length is shorter and the preservation of the local polarization degree is improved.
This monotonic dependence is confirmed by the results of $\eta=0.01$ [see Fig.~\ref{fig:Figures_chapter5_fig_2023Gong_PRL_slingshot}(c)(d)], where electrons experience stochastic acceleration and photon emission is isotropic in the angular space.
However, for $\eta=0.4$ [see Fig.~\ref{fig:Figures_chapter5_fig_2023Gong_PRL_slingshot}(c)(d)], the averaged polarization degree $\left<\mathcal{Q}\right>$ versus $\omega_{ph}$ exhibits non-monotonic dependence, with a polarization dip $\Delta \mathcal{Q}_\omega\approx 4.5\%$  and a bandwidth ratio $\mathcal{B}_\omega\equiv \omega_{ph}^m/\omega_{ph}^*\sim 10^3$, contradictory to the aforementioned monotonic dependence. Here, $\omega_{ph}^*$ is the local maximum point of the function $\left<\mathcal{Q}\right>$ vs $\omega_{ph}$ [see Fig.~\ref{fig:Figures_chapter5_fig_2023Gong_PRL_slingshot}(c)].
Using the correlation between the work contribution and the non-monotonic dependence of photon polarization, Gong \textit{et al.}~\cite{Gong2023_slingshot} identified the mechanism of electron slingshot acceleration.

Another recent study by Gong \textit{et al.}~\cite{gong2025spin} reported a understanding of anomalous gamma-ray photon polarization in radiation-dominated magnetic reconnection. Using particle-in-cell simulations that account for radiation reaction, spin dynamics, and quantum polarization effects, Gong \textit{et al.} found the formation of spin-polarized condensed plasmoids. Electrons in the reconnection layer contract in phase space toward a spiral attractor induced by radiation damping, leading to the emergence of compact, high-density plasmoids with spin alignment along the local magnetic field. Compared with conventional plasmoids, these structures demonstrate much stronger density compression. More importantly, their radiation displays an unexpected linear polarization: the emitted gamma-ray photons are polarized perpendicular to the electron motion plane, in sharp contrast to classical synchrotron predictions. The study identifies spin-flip transitions as the key mechanism that reshapes angular momentum transfer and produces this anomalous polarization. This finding provides not only theoretical evidence for condensed spin-ordered plasmoids in magnetic reconnection, but also a new perspective for explaining puzzling polarization signals detected from extreme astrophysical environments such as magnetars~\cite{sironi2014relativistic,hoshino2023energy,bacchini2024collisionless}.

\subsection{Vortex gamma-ray photons}

Vortex gamma-ray photons are high-energy photons that carry intrinsic orbital angular momentum (OAM) in addition to their intrinsic spin angular momentum~\cite{taira2017gamma,allen1992orbital,knyazev2018beams,ivanov2022promises,Wang2022_oam,Bu2024_vortex}. Characterized by a helical wavefront structure, these photons exhibit a quantized twist, allowing them to have unique properties that differ from ordinary gamma rays. 
Vortex photons have been successfully generated across the visible to X-ray range using techniques such as optical mode conversion, high harmonic generation, and coherent radiation from helical undulators or laser facilities~\cite{shen2019optical,terhalle2011generation,gariepy2014creating,hemsing2013coherent,Zhou2025_x-ray_vortices}.
As the new degree of freedom of information is carried by the vortex state, vortex gamma-ray photons have broad applications in various fields, including astrophysics, strong-field laser physics, particle physics, and nuclear physics~\cite{afanasev2018radiative,schulz2019modification,afanasev2021recoil,bu2021twisted,lu2023manipulation}.
However, the generation of vortex gamma photons remains a big challenge.

With the rapid progress of ultraintense laser techniques, most studies utilize Compton backscattering to obtain high-energy and OAM gamma-ray photons~\cite{liu2016generation,taira2017gamma,taira2018gamma,gong2018brilliant,zhu2018generation,chen2018gamma,feng2019emission,wang2020generation,liu2020vortex,hu2021attosecond,zhang2023generation,younis2022generation}. %\textcolor{red}{Nevertheless, most of these studies have only considered vortex beams with collective OAM in laser-plasma interactions, the intrinsic OAM carried by gamma-ray photons is still unclear.}
However, the produced OAM-dominated beams can be described semiclassically as ensembles of particles moving at an angle that carry classical OAM about the beam axis; a rotating cloud whose large mechanical OAM arises purely as a collective classical effect. 
In contrast, OAM in vortex states is an intrinsic property of individual particles, not a collective effect.
From the classical electro-magnetic field radiation of energetic electron beams colliding with the vortex laser field, it is found that the accumulative field structure shows vortex features~\cite{petrillo2016compton}. Inherent OAMs carried by gamma-photons have been shown in linear Compton scattering between energetic electrons and vortex laser photons, in the framework of quantum electrodynamics (QED)~\cite{jentschura2011generation}. 

In the nonlinear regime, the QED scattering theory including the quantum number of angular momentum has been considered to characterize multiphoton absorption and the process of angular momentum transfer in the generation of vortex gamma-ray photons. For instance, Bu \textit{et al.} \cite{Bu2024_vortex} showed that in Compton scattering of ultrarelativistic electrons in a CP laser pulse, absorption of spin-angular-momenta of multiple CP laser photons leads to efficient generation of twisted gamma photons in the weak linear regime and also OAM electrons in the strong linear case. Ababekri \textit{et al.}~\cite{ababekri2024vortex} explored the generation of vortex gamma photons in similar setups. They discuss the vortex phase structure of the scattering matrix element, the transformation of the vortex phase into the emitted photon, and the radiation rate for vortex gamma photons. 
Guo \textit{et al.}~\cite{guo2024generation} investigated the production of gamma-ray photons with large intrinsic OAM via nonlinear Compton scattering. Gamma-ray photons carrying controllable OAM quantum numbers from tens to thousands of units are generated when employing vortex electrons at moderate laser intensities.  
Jiang \textit{et al.}~\cite{jiang2024_} studied the generation of vortex gamma photons with controllable spin and orbital angular momenta via nonlinear Compton scattering of two-color counter-rotating CP laser fields. They find that the polarization and vortex charge of the emitted gamma photons can be controlled by tuning the relative intensity ratio of the two-color CP laser fields.
Alternatively, vortex gamma photons can also be efficiently generated through Bremsstrahlung, whether from initially twisted electrons~\cite{wang2022triple} or through spin-to-orbital coupling with spin-polarized electrons~\cite{wang2022finite}.

Utilizing quantum radiation theory could lead to a better understanding of the properties of the emitted vortex gamma-ray photons, which potentially opens new opportunities in hadron, atomic, nuclear, particle, and high-energy physics. Further detailed discussion of the promises and challenges of high-energy vortex photons and other vortex particles can also be found in relevant review papers, e.g. Bliokh \textit{et al.}~\cite{bliokh2017theory}, Lloyd \textit{et al.}~\cite{lloyd2017electron}, Larocque \textit{et al.}~\cite{larocque2018twisted}, and Ivanov~\cite{ivanov2022promises}.

\section{Summary and future prospects} \label{sec:summary}

Over the last few years, the area of laser-plasma based acceleration methods for spin-polarized particle beams has significantly evolved. This is in part due to the advances in general plasma-based accelerator research as well as the recent studies into the generation of polarized particle sources.

Up to now, most of the proposed mechanisms rely on pre-polarized targets, which come with certain restrictions on the maximum attainable density and size. Only a few results suggest the utilization of \textit{in-situ} setups as in the case of polarized electron beams \cite{Nie2021_in_situ, Nie2022_single_species}. In the regime of QED physics, the methods rely on radiative polarization \cite{Seipt2018_radiative, Tang2021_radiative_dynamics}.
For pre-polarized targets, it has become obvious in most of the theoretical studies that the injection phase of the particles into the accelerating structure is the most crucial for the preservation of polarization. The subsequent acceleration phase, where $\gamma \gg 1$, was shown to only marginally contribute to depolarization.
The recent proof-of-principle results with ${}^3$He give hope that laser-plasma based polarized targets are a feasible approach \cite{Zheng2023_evidence_he3}.

The next logical step in this field will be experimental verification of the proposed setups for electron and ion acceleration, since the amount of theoretical studies currently outweighs experimental results. The generation of polarized positrons and gamma quanta will further rely on any advances in general accelerator research, as high-intensity laser pulses and/or high-energy electron beams are necessary for many of these schemes.

These might become feasible with near-future laser facilities like ELI \cite{ELI}, XCELS \cite{XCELS} or SULF/SEL at SIOM \cite{SIOM}.

Laser–plasma acceleration naturally produces ultrashort pulses (on the order of femtoseconds or picoseconds) with extremely high peak currents in the kiloampere range. These values exceed conventional polarized sources, such as GaAs photocathodes, by three to four orders of magnitude in peak current. The combination of femtosecond/picosecond duration and kiloampere currents provides new opportunities for nuclear and particle physics.

For instance, polarized deep inelastic scattering (DIS) experiments have historically been limited by the achievable beam current from conventional polarized injectors \cite{Burkardt2010_scattering_nucleons, Ageev2005_DIS_deuteron}. A kiloampere-class polarized beam would enable instantaneous luminosities orders of magnitude higher, significantly improving statistical precision in measurements of the nucleon spin structure.
In polarized proton–deuteron or proton–nucleus scattering, high-current polarized injectors would allow precise mapping of spin-dependent nuclear forces \cite{Glashausser1979_nuclear}. The short pulse duration additionally provides time-resolved probes of nuclear excitations, something not possible with CW polarized beams.

Laser-driven polarized beams may also impact fusion research, where spin alignment can enhance nuclear reaction cross sections. In particular, in the context of proton-boron (p-${}^{11}$B) fusion, polarized protons accelerated in a laser-plasma environment could modify the angular distribution of alpha particles. The reaction yield can be enhanced if the target is also polarized. The ultrashort pulse duration ensures synchronization with compression or ignition stages, potentially enabling polarized fast-ignition schemes in laser-driven fusion systems.

Perhaps the most immediate application of laser-driven polarized particles is as injectors into conventional accelerator facilities. Conventional polarized sources, such as strained-GaAs photocathodes, are constrained to average currents in the milliampere regime. In contrast, laser-plasma injectors can deliver kiloampere-scale polarized bunches that can be captured and accelerated further in RF linacs or storage rings. This approach provides access to high-energy, high-brightness polarized beams well beyond the limits of conventional injectors, supporting future polarized electron-ion colliders (EICs) at higher instantaneous luminosity for spin-structure studies.

\ack
Z.G. would like to thank Karen Z. Hatsagortsyan, Christoph H. Keitel, Matteo Tamburini, and Xiaofei Shen for fruitful discussions. L. J. acknowledges the support by the National Science Foundation of China (No. 12388102) and the Strategic Priority Research Program of the Chinese Academy of Sciences (No. XDB0890000).

This work has been funded in parts by the DFG (projects PU 213/9-1 and BU 2227/5-1) and BMBF (project 05P24PF1).
The work of M.B. has been carried out in the framework of the JuSPARC (Jülich Short-Pulse Particle and Radiation Center \cite{Buescher2020_jusparc}).

%\section*{References}
%\bibliographystyle{iopart-num}
%\bibliography{spin_ropp.bib}

\input{polarized_beams_rep_prog_phys.bbl}

\end{document}

%% file: polarized_beams_rep_prog_phys.bbl
\providecommand{\newblock}{}